\documentclass[useAMS,usenatbib,usegraphicx]{mn2e}
\usepackage{url}
\usepackage{comment}
\usepackage{color}

\def\Lya{Ly$\alpha$}
\def\Ha{H$\alpha$}
\def\Hb{H$\beta$}
\def\OII{[O\,{\sc ii}]}
\def\OIII{[O\,{\sc iii}]}
\def\NII{[N\,{\sc ii}]}

\def\NeIII{[Ne\,{\sc iii}]}

\def\fesc{$f_{\rm esc}$}

\def\cms{cm\,s$^{-1}$}

\def\Oabundance{$12+\log ({\rm O/H})$}
\def\HI{H\,{\sc i}}
\def\HII{H\,{\sc ii}}
\def\qion{$q_{\rm ion}$}
\def\NHI{$N_{\rm HI}$}
\def\Stromgren{Str\"{o}mgren}

\def\aj{AJ}
\def\apj{ApJ}
\def\apjs{ApJS}
\def\apjl{ApJL}
\def\aap{A\&A}	
\def\mnras{MNRAS}
\def\pasj{PASJ}
\def\pasp{PASP}

\def\nat{Nature}

\title[Ionization State of ISM in Galaxies]
{Ionization State of Inter-Stellar Medium in Galaxies: 
Evolution, SFR-$M_{\star}$-$Z$ Dependence, and Ionizing Photon Escape}

\author[K. Nakajima \& M. Ouchi]{
Kimihiko Nakajima$^{1,3,4}$\thanks{E-mail: kimihiko.nakajima@nao.ac.jp}
and
Masami Ouchi$^{2,3}$\\
$^1$Department of Astronomy, Graduate School of Science,
    The University of Tokyo, 7-3-1 Hongo, Bunkyo-ku,
    Tokyo 113-0033, Japan\\
$^2$Institute for Cosmic Ray Research,
    The University of Tokyo, 5-1-5 Kashiwanoha, Kashiwa,
    Chiba 277-8582, Japan\\
$^3$Kavli Institute for the Physics and Mathematics of
    the Universe (WPI),
    The University of Tokyo, 5-1-5 Kashiwanoha, Kashiwa, \\
    Chiba 277-8583, Japan\\
$^4$National Astronomical Observatory of Japan, 
    2-21-1 Osawa, Mitaka, 
    Tokyo 181-8588, Japan\\
}

\begin{document}

\date{Accepted for publication in MNRAS}

\pagerange{\pageref{firstpage}--\pageref{lastpage}} \pubyear{2014}

\maketitle


\begin{abstract}

We present a systematic study for 
ionization state of inter-stellar medium in galaxies 
at $z=0-3$ with $\sim 140,000$ SDSS galaxies and $108$ intermediate to high 
redshift galaxies from the literature, using an ionization-parameter 
sensitive line ratio of \OIII$\lambda5007$$/$\OII$\lambda 3727$ and
photoionization models.
We confirm that $z\sim 2-3$ galaxies show an \OIII$/$\OII\ ratio
significantly higher than 
a typical star-forming galaxy of SDSS 
by a factor of $\ga 10$, 
and the photoionization models reveal that these high-$z$ galaxies have 
an ionization parameter of $\log($\qion$/$\cms$)\sim 7.6-9.0$, a factor 
of $\sim 4-10$ higher than local galaxies.
For galaxies at any redshift, 
we identify a correlation between the 
\OIII$/$\OII\ ratio and galaxy global properties of star-formation rate 
(SFR), stellar mass ($M_{\star}$), and metallicity ($Z$). We extend the 
fundamental metallicity relation (FMR) 
and develop the fundamental ionization relation (FIR), 
a four-dimensional relation of ionization 
parameter, SFR, $M_{\star}$, and $Z$. The intermediate and high-$z$ 
galaxies up to $z\simeq 3$ follow the FIR defined with the local 
galaxies, in contrast with the FMR whose possible evolution from 
$z\sim 2$ to $3$ is reported. We find that the FMR evolution of 
$z\sim 2-3$ appears, if one omits ionization parameter differences, 
and that the FMR evolution does not exist for an average metallicity 
solution of $z\sim 3$ galaxies with a high-ionization parameter.
Interestingly, all of two
local Lyman-continuum emitting galaxies (LyC leakers) 
have a high \OIII$/$\OII\ ratio, indicating a positive correlation
between \OIII$/$\OII\ and ionizing photon escape fraction (\fesc), 
which is successfully explained by our photoionization models. Because 
\OIII$/$\OII\ ratios of $z\sim 2-3$ galaxies, especially Ly$\alpha$ 
emitters (LAEs), are comparable to, or higher than, those of the local 
LyC leakers, these high-$z$ galaxies are candidates of Lyman-continuum 
emitting objects. A strong Ly$\alpha$ emission can coexist with a large 
\fesc\ of $\la 0.8$, and the increasing fraction of LAEs towards 
high-$z$ reconciles the picture of cosmic reionization whose major 
ionizing sources are faint galaxies having {\it intrinsically}-bright
\Lya\ emission.

\end{abstract}



\section{Introduction} \label{sec:introduction}

For galaxy formation studies, it is essential to understand physical 
conditions of inter-stellar medium (ISM) of galaxies. Physical properties 
of hot ISM are characterized by gas-phase metallicity and ionization 
parameter.
A gas-phase metallicity is a record of galaxy's star formation history 
and gas infall$+$outflow, and is commonly defined by oxygen abundance, 
\Oabundance.
Metallicity estimates can be made with metal lines divided by a hydrogen
recombination line, such as
(\OIII$\lambda\lambda 5007,4959$$+$\OII$\lambda 3727$)$/$\Hb\
($R23$-index; e.g., \citealt{pagel1979,KD2002}) and
\NII$\lambda 6584$$/$\Ha\ ratio
($N2$-index; e.g., \citealt{storchi-bergmann1994,KD2002,PP2004}).
An ionization parameter, \qion, is sensitive to the degree of excitation
of an \HII-region. It is defined as the ratio of mean ionizing photon flux 
to mean atom density, and its dimensionless form, $U=q_{\rm ion}/c$,
is also used.
Ionization parameters are estimated with emission lines of different 
ionization stages of the same element, such as
\OIII$\lambda 5007$$/$\OII$\lambda 3727$ ratio (e.g., \citealt{KD2002}).

Metallicities of galaxies have been well studied in the local and 
high-$z$ universe in the diagram of stellar-mass vs. metallicity 
($M_{\star}$--$Z$) relation. In the local universe, a tight 
$M_{\star}$--$Z$ relation is recognized, where a metallicity increases 
with stellar mass (e.g., \citealt{tremonti2004}). At high-$z$, similar 
$M_{\star}$--$Z$ relations are reported at $z=0.5-3.5$ (e.g.
\citealt{savaglio2005,erb2006,maiolino2008,mannucci2009,zahid2011,yabe2012}).
These studies find the evolution of $M_{\star}$--$Z$ relation; 
a metallicity decreases towards high-$z$ at a given $M_{\star}$.
Recently, \citet{mannucci2010} and \citet{lara-lopez2010} identify 
that the $M_{\star}$--$Z$ relation depends on star formation rate (SFR),
which is initially suggested by \citet{ellison2008}, and propose the 
fundamental metallicity relation (FMR) that forms a single plane
in the stellar-mass, metallicity, and SFR space. The FMR holds over 
$z\simeq 0-2.5$ with no evolution. This would indicate that there are 
physical processes commonly found in local and high-$z$ galaxies.
Many follow-up studies test the universality of the FMR over the wide 
range of stellar mass, SFR, and redshifts
(e.g, \citealt{mannucci2011,richard2011,nakajima2012,niino2012,%
yabe2012,cresci2012,wuyts2012,christensen2012a,belli2013,lara-lopez2013}).
Although galaxies at $z\la 2.5$ follow the FMR, galaxies at $z\ga 3$ 
appear to fall below the FMR \citep{mannucci2010}. \citet{mannucci2010} 
claim that this disagreement suggests that there are physical 
mechanisms different from those of low-z galaxies at $z\sim 3$. 
However, this is still an open question.

In contrast to metallicities, ionization parameters of galaxies
are poorly understood. Recently, \citet{nakajima2013} use a diagram 
of \OIII$/$\OII\ ratio and $R23$-index to study ionization parameters 
and metallicities for galaxies at $z=0-3$, focusing on the properties 
of Ly$\alpha$ emitters (LAEs) at $z\sim 2$.
\citet{nakajima2013} report that a typical ionization parameter 
increases towards high-$z$, and suggest the evolution of ionization 
parameter (see also, \citealt{lilly2003,hainline2009,richard2011}). 
However, until recently, statistical studies for ionization parameters
do not cover
(i) physical relations between ionization parameter and galaxy global
properties, such as stellar mass and SFR,
and
(ii) evolution of the physical relations with ionization parameter.
\citet{dopita2006a} present that in the local universe an ionization 
parameter seems to gradually decrease with $M_{\star}$, but do not 
address the issue about evolution of this trend
(see also \citealt{brinchmann2008}).

There is another diagnostics of ionization parameter, using a plot 
of \NII$/$\Ha\ and \OIII$/$\Hb\ ratios that is referred to as the 
BPT diagram \citep{baldwin1981}.
Local star-forming galaxies form a tight sequence on the BPT diagram 
(e.g., \citealt{kauffmann2003a}). In contrast, high-$z$ galaxies depart 
from the local star-forming sequence towards higher \OIII$/$\Hb\ 
ratio (e.g., \citealt{shapley2005,erb2006,liu2008,hainline2009,%
finkelstein2009,yabe2012}). 
Similarly, recent spectroscopic surveys report that high-$z$ 
galaxies on average have \OIII$/$\Hb\ ratios much higher than 
local similar mass galaxies
\citep{cullen2014,holden2014}.
One interpretation of that departure is 
evolution of ionization parameter
(e.g., \citealt{brinchmann2008,kewley2013a,kewley2013b}).
\citet{kewley2013a} have compared high-$z$ galaxy line ratios and 
theoretical predictions to investigate evolution of the star-forming
sequence on the BPT diagram, and claimed that high-$z$ galaxies would 
have a larger ionization parameter, a higher electron density,
and/or a harder ionizing radiation field 
(see also \citealt{brinchmann2008,kewley2013b}).
\citet{shirazi2013} have also argued a possibility that the 
typical gas density of \HII-regions would be much higher in  
high-$z$ galaxies.

Ionization parameters are also important for accurate estimates of 
metallicity from nebular emission lines. Metallicity estimation 
methods based on local empirical relations 
(e.g., \citealt{maiolino2008}) are often applied to high-$z$ galaxies.
These methods implicitly assume ionization parameters of typical local 
galaxies. Since various studies suggest that ionization parameter 
evolves with redshift, the estimated metallicities from local 
empirical relations would be biased.

Recently, a relationship between ionization parameter and escape 
fraction of ionizing photons (\fesc) and its importance for cosmic 
reionization are discussed. \citet{nakajima2013} have pointed out
a possibility that \fesc\ positively correlates with ionization 
parameter, since an optically-thin nebula could show a high 
\OIII$/$\OII\ ratio
(e.g., \citealt{giammanco2005,brinchmann2008,kewley2013b}).
If this correlation of ionization parameter and \fesc\ is established,
ionization parameters allow us to investigate ionizing photon escapes
that cannot be directly observed for galaxies at the epoch of 
reionization ($z>6$; \citealt{fan2006,dunkley2009}).
Studies of galaxies at $z\sim 6-7$ have identified a tension between 
the ionizing photon production rate and cosmic reionization 
(e.g., \citealt{ouchi2009,robertson2010}). A moderately high average 
\fesc\ of $\ga 0.2$ would be required for galaxies in the early 
universe to keep the inter-galactic medium (IGM) ionized.
\citet{nakajima2013} have suggested that LAEs exhibit higher ionization 
parameters than other high-$z$ galaxy populations such as Lyman-break 
galaxies (LBGs). Because observations reveal that a fraction of LAEs to
star-forming galaxies of LBGs increases with redshift
(e.g., \citealt{ouchi2008,vanzella2009,stark2010,stark2011}), more LAEs 
with a high ionization 
parameter, i.e., a high \fesc, would ease the tension of the deficit of 
ionizing photons at the early universe.
However, the claim of \citet{nakajima2013} is made only with two LAEs, 
and it is unclear whether this is a general feature of LAEs.
One needs to
(i) investigate the trend that LAEs typically show an ionization 
parameter higher than LBGs, and
(ii) understand the physical reason of the correlation between 
ionization parameter and \fesc\ with theoretical models.
Although predicting \fesc\ is challenging by using simulations or 
theoretical models due to the large uncertainties in the 
assumptions about the ISM and the limited resolutions 
(e.g., \citealt{rahmati2013}), we investigate the correlation
with photoionization models assuming a simple, homogeneous gas cloud 
as the first step.

This paper is organized as follows. We describe our samples of 
galaxies in Section \ref{sec:samples}, and test the main sample of SDSS 
in Section \ref{sec:testing_sample}.
In Section \ref{sec:o3o2r23}, we show all of our galaxies in the 
\OIII$/$\OII\ vs. $R23$-index diagram, and investigate galaxy 
properties in the diagram with the aid of photoionization models.
We then clarify the dependencies of ionization parameter on galaxy
global properties of stellar mass, SFR, and metallicities.
We introduce a new relation named fundamental ionization relation 
(FIR) that is described in the four-dimensional parameter space of
\OIII$/$\OII\ ratio, $R23$-index, stellar mass, and SFR.
We discuss these observational results in Section \ref{sec:discussion},
addressing the issues of
(i) the evolution of ionization parameter,
(ii) the FMR evolution at $z\ga 3$,
(iii) the relations between \OIII$/$\OII\ ratio, $R23$-index,
stellar mass, and SFR, and
(iv) the correlation between ionization parameter and \fesc.
Finally, we summarize our results in Section \ref{sec:summary}.
Throughout this paper, we estimate stellar masses and SFRs based on a 
\citet{chabrier2003} initial mass function (IMF), and
assume an \citet{asplund2009} solar metallicity, where
$\log(Z/Z_{\odot})=$ \Oabundance\,$-8.69$.
We adopt a concordance $\Lambda$CDM cosmology with
$(\Omega_m,\, \Omega_{\Lambda},\, H_0)=
(0.3,\, 0.7,\, 70\,{\rm km}\,{\rm s}^{-1}\,{\rm Mpc}^{-1})$.

\section{Galaxy Samples} \label{sec:samples}

\subsection{Local Galaxies} \label{ssec:sample_z0}

\subsubsection{SDSS galaxies} \label{sssec:sample_z0_sdss}

We make an SDSS sample from the SDSS Data Release 7 MPA/JHU catalog%
\footnote{
\url{http://www.mpa-garching.mpg.de/SDSS/DR7/}
}
including emission line properties and stellar masses
(\citealt{kauffmann2003b,brinchmann2004,salim2007}).
We select galaxies with the criteria similar to those adopted by 
\citet{mannucci2010};
(1) $0.07<z<0.30$,
(2) a signal-to-noise ratio (S/N) of \Ha\ $>25$,
(3) $A_V<2.5$,
(4) \Ha$/$\Hb\ $>2.5$,
(5) active galactic nuclei (AGN) are removed based on the BPT diagram
\citep{baldwin1981} with the classification of \citet{kauffmann2003a}%
\footnote{
Although \citet{brinchmann2004} recommend that requiring S/N $>3$ 
in all four BPT diagram lines is useful, we do not adopt the criteria 
because the requirement could bias the sample regarding metallicity.
Indeed, the criteria of S/N ratio of \Ha\ $>25$ is strong enough to 
remove lots of galaxies with low-S/N emission lines in the sample. 
By imposing S/N ratios of all four BPT diagram lines $>3$, the sample
size decreases to $\sim 120,000$. Nevertheless, we have checked that 
our main results do not depend significantly on the criteria. 
In order to fairly compare with the FMR of \citet{mannucci2010}, 
we do not impose the requirement of S/N ratios for the BPT diagram.
},
(6) metallicities estimated from the $N2$- and $R23$-index
\citep{maiolino2008} agree within $0.25$\,dex.
We adopt the average value of the two metallicities%
\footnote{
These metallicities are only used for the consistency check
of the FMR of \citet{mannucci2010} 
(Section \ref{sec:testing_sample}).
}%
, and use total 
stellar masses listed in the updated MPA/JHU stellar mass catalog%
\footnote{
\url{http://home.strw.leidenuniv.nl/~jarle/SDSS/}
}.
All the emission line fluxes are corrected for extinction using 
\Ha$/$\Hb\ and the extinction curve given by \citet{cardelli1989}.
SFRs are estimated from the \Ha\ luminosities by using the 
\citet{kennicutt1998}'s relation, 
and (total) stellar masses from the broadband photometry.
All of the SFRs and stellar masses are corrected to the 
\citet{chabrier2003} IMF.
We remove the same objects in the catalog that are found by the 
MPA/JHU group. 
We thus obtain $136,067$ SDSS galaxies.

In this sample, we highlight two populations of galaxies shown below.
\begin{itemize}
  \item Green Pea galaxies (GPs; \citealt{cardamone2009}) that are 
    local star-forming galaxies with a very strong \OIII\ emission 
    line whose EW is $\sim 1000$\,\AA. They are found in the SDSS
    spectroscopic sample of galaxies with green colors in the SDSS 
    $gri$ composite images, where the strong \OIII\ emission fall
    in the $r$ band. Because the strong \OIII\ emission has to be
    redshifted to the bandpass of $r$, the redshift of GPs ranges
    from $0.112$ to $0.360$.
    We use $80$ GPs given by \citet{cardamone2009} that are free 
    from AGN contamination. We cross-match the $80$ GPs to our 
    SDSS galaxies, and identify $51$ GPs. To include GPs at 
    $z>0.30$, we perform the same cross-matching to SDSS galaxies
    that do not meet the redshift criterion of (1), and obtain 
    additional $15$ GPs at $z>0.3$. We use a total of $66=(51+15)$ 
    GPs in our analysis.
  \item Lyman Break Analogs (LBAs; \citealt{heckman2005,hoopes2007,%
    basu-zych2007,overzier2008,overzier2009}) that are super-compact
    UV-luminous local galaxies whose characteristics are similar to 
    those of high-$z$ LBGs. The UV imaging survey performed by the 
    {\it Galaxy Evolution Explorer (GALEX)} is used to select 
    star-forming galaxies at $0.1<z<0.3$ with a high luminosity of 
    $L_{\rm FUV}>10^{10.3}\,L_{\odot}$ and a high far UV surface 
    brightness of $I_{\rm FUV}>10^{9}\,L_{\odot}\,{\rm kpc}^{-2}$.
    We compile catalogs of \citet{hoopes2007}, \citet{basu-zych2007}, 
    and \citet{overzier2009}, and obtain $51$ LBAs. Cross-matching 
    them to our SDSS galaxies and removing AGN candidates, 
    we make an LBA sample composed of $37$ LBAs.
\end{itemize}
We remove sources in our GP and LBA samples from our SDSS galaxies,
and define a sample of SDSS galaxies with no GPs and LBAs as 
SDSS sample.

\begin{figure*}
  \centerline{
    \includegraphics[width=0.98\textwidth]{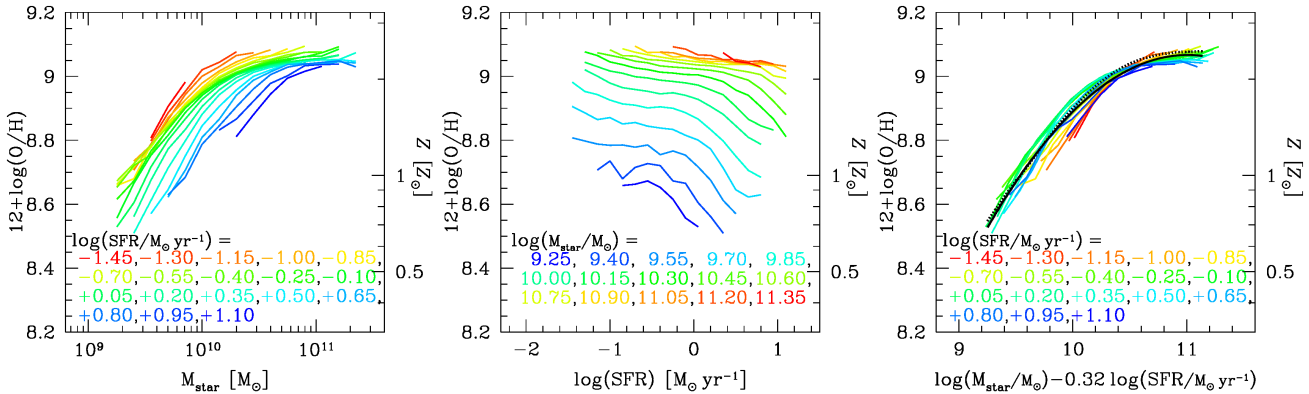}
  }
  \caption{
    {\em Left:} The mass-metallicity relations of galaxies from our SDSS 
    sample. Each line denotes a SFR subsample from our SDSS sample whose
    SFR is presented at the bottom of the panel. 
    {\em Center:} The SFR-metallicity relation given by our SDSS sample.
    The lines represent the relations for subsamples with $M_{\star}$ 
    values shown in the legend.
    {\em Right:} Same as the left panel, but for the 
    $\mu_{0.32}$-metallicity relation, a two-dimensional view of the FMR. 
    The color code is the same as the one in the left panel. The black 
    solid curve is the quadratic best-fit of our SDSS sample, while the 
    black dotted curve is the original FMR obtained by 
    \citet{mannucci2010} and \citet{mannucci2011}. The black solid and 
    dotted lines are almost identical. Thus, our SDSS sample reproduces 
    the original FMR.
  }
  \label{fig:fmr_Man10}
\end{figure*}

\subsubsection{LyC leakers}
\label{sssec:sample_z0_LyC}

There are two local star-forming galaxies emitting Lyman-continuum 
(LyC) radiation, Haro 11 \citep{bergvall2006,leitet2011} and 
Tol 1247-232 \citep{leitet2013}. 
They are initially selected as blue compact galaxies or \HII\ galaxies,
and found to leak LyC radiation from the Far Ultraviolet Spectroscopic 
Explorer (FUSE) data.
In order to study the relation between 
ionization state and ionizing photon escape, we investigate the two LyC 
leaking galaxies. 
Hereafter we refer to these two galaxies as LyC leakers.
Emission line fluxes for Haro 11 and Tol 1247-232 are obtained from
\citet{BO2002} and \citet{terlevich1993}, respectively, and stellar 
masses from \citet{leitet2013}.
Their SFRs are derived in the same manner as those of our SDSS galaxies
(Section \ref{sssec:sample_z0_sdss}).
Neither of them is contaminated by AGN \citep{leitet2013}.
We note that Haro 11 can be classified as an LBA \citep{heckman2011}.

\subsection{Intermediate and High Redshift galaxies} 
\label{ssec:sample_highz}

We make samples of intermediate- and high-$z$ galaxies that have firm 
estimates of stellar mass and spectroscopic measurements of nebular 
emission lines (hydrogen Balmer lines, \OII, \OIII). We particularly 
include strongly lensed galaxies, BX/BM galaxies, LBGs, and LAEs.
\begin{itemize}
  \item Intermediate redshift galaxy sample ($z=0.5-1.5$) is composed 
    of $24$ galaxies from the Gemini Deep Deep Survey (GDDS) and $15$ 
    galaxies from the Canada-France Redshift Survey (CFRS) 
    \citep{savaglio2005,maier2005}. Moreover, we add $26$ low-mass 
    galaxies recently reported by \citet{henry2013}, and $4$ strongly 
    lensed galaxies \citep{christensen2012a,queyrel2009} to the sample.
    In total, the intermediate-$z$ sample contains $69$ galaxies.
  \item High redshift galaxy sample ($z\sim 2-3$) consists of $39$ 
    galaxies in total, $21$ strongly lensed galaxies \citep{pettini2001,%
    fosbury2003,richard2011,rigby2011,christensen2012a,wuyts2012,%
    belli2013}, two LAEs \citep{nakajima2013}, one metal-poor BX galaxy 
    \citep{erb2010}, and $15$ LBGs from the AMAZE and LSD projects
    \citep{maiolino2008,mannucci2009}.
\end{itemize}
We use a total of $108=(69+39)$ intermediate- and high-$z$ galaxies
in our analysis.

We divide the high-$z$ galaxy sample into two subsamples, LAE and LBG 
samples, to investigate the relationship between ionization state and 
\Lya\ photon escape.
For high-$z$ galaxies with a \Lya\ flux measurement, we classify 
galaxies with EW$_{\rm rest}$ (\Lya) $>20$\,\AA\ as LAE; Lynx arc
\citep{fosbury2003}, BX418 \citep{erb2010}, Abell 1689 31.1 arc, and 
SMACS J2031 arc \citep{christensen2012a,christensen2012b}. We include 
the two narrow-band selected LAEs, CDFS-3865 and COSMOS-30679 
\citep{nakajima2013}, to the LAE sample.
For high-$z$ galaxies with EW$_{\rm rest}$ (\Lya) $<20$\,\AA\  
or no \Lya\ flux measurement, we regard these 
galaxies as LBGs for simplification, 
given the observational fact that continuum-selected galaxies 
on average have an EW of \Lya\ $\sim 0$\,\AA\ 
(e.g., \citealt{shapley2003,kornei2010}).
Indeed, most of the $z\sim 3$ galaxies in the sample of 
\citet{maiolino2008} are originally spectroscopically-confirmed 
based on rest-frame UV absorption lines, not \Lya\ emission line
(see also \citealt{vanzella2006}).

We derive physical quantities for these intermediate- and high-$z$ 
galaxies with the same procedures as those for our SDSS galaxies. 
Emission line fluxes are corrected for extinction with \Ha$/$\Hb\ and 
the extinction curve given by \citet{calzetti2000}. For galaxies with 
no \Ha\ or \Hb\ measurement, we correct for extinction estimated from 
spectral energy distribution (SED) fitting, assuming that the amount of 
extinction is the same in continuum and emission lines. For galaxies 
neither with \Ha$/$\Hb\ nor dust extinction from SED fitting%
\footnote{
These galaxies are $z\sim 1$ GDDS objects.
},
we adopt dust extinction inferred from stellar mass based on the 
empirical relation \citep{gilbank2010}. SFRs are estimated from a 
hydrogen recombination line of \Ha\ or \Hb\ corrected for dust 
extinction with the \citet{kennicutt1998}'s relation.
If \OIII$\lambda 4959$ fluxes are not available in the literature%
\footnote{
There are $25$, $10$, and $9$ galaxies at $z\sim 1$, $2$, and $3$,
respectively.
}, we assume a value of $0.28$ for the \OIII$\lambda4959/5007$ ratio
\citep{richard2011}. We have confirmed that this assumption does
not affect our results significantly. We estimate intrinsic SFRs and 
stellar masses, using lens magnification factors for the strongly 
lensed galaxies. We place the lower (upper) limits on the SFR and 
stellar mass of SGAS1527 (SGAS1226) that only has the upper (lower)
limit of the magnification factor \citep{wuyts2012}.
We adopt the lower limit of stellar mass given by 
\citet{villar-martin2004} for Lynx arc \citep{fosbury2003}.
For high-$z$ galaxies with no \OII\ detections
\citep{fosbury2003,erb2010,richard2011},
we apply $3\sigma$ upper-limit fluxes of \OII\ for the lower limits
of \OIII$/$\OII, while we approximate $R23$-indices under the 
assumption of no significant \OII\ flux. Using the BPT diagram, we 
have identified no obvious AGN activity in the high-$z$ galaxies.
\\

We note that all the properties for the local, 
intermediate-$z$, and high-$z$ galaxies used in this paper are 
derived in a consistent manner, and that they can be fairly
compared to each other.

\begin{figure*}
  \centerline{
    \includegraphics[width=0.75\textwidth]{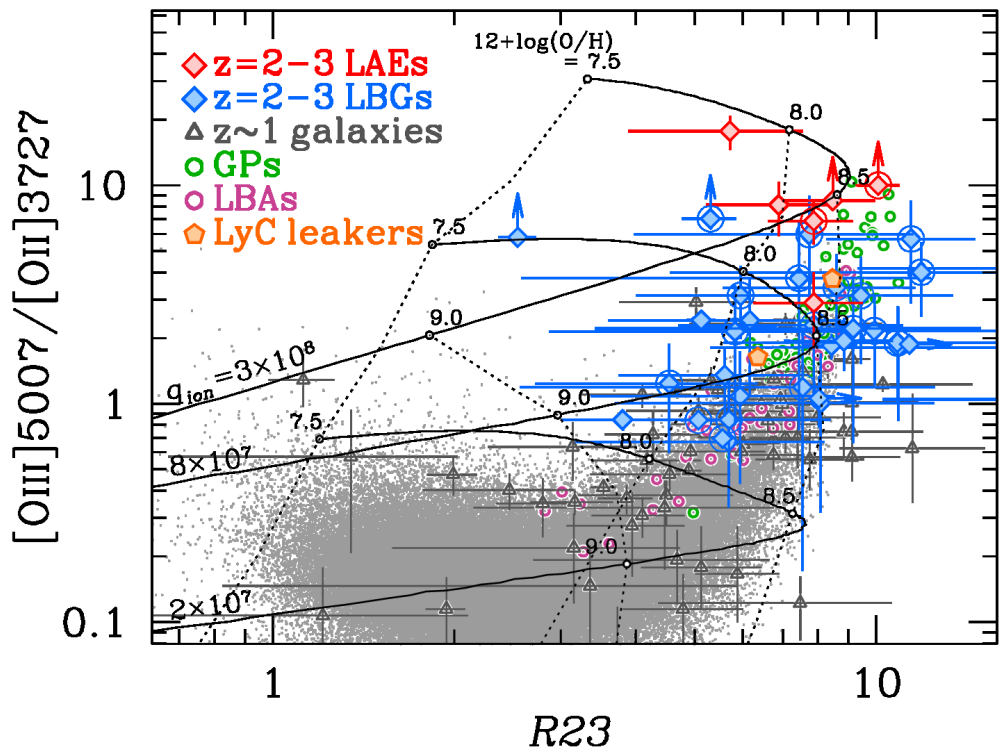}
  }
  \caption{
    The relation between \OIII$/$\OII\ and $R23$-index.
    The gray dots are the SDSS galaxies. The red and blue diamonds 
    represent $z=2-3$ LAEs and LBGs, respectively.
    Specifically, the diamonds with a circle denote galaxies at $z>3$.
    The gray triangles are $z\sim 1$ star-forming galaxies. The green
    and purple circles show GPs and LBAs, respectively, and the orange
    pentagons are LyC leakers in the local universe.
    The black solid curves present photoionization model tracks in the 
    range of \Oabundance$=7.5-9.5$ for \qion\ $=3\times 10^8$, 
    $8\times 10^7$, and $2\times 10^7$\,cm\,s$^{-1}$ \citep{KD2002}.
    The numbers with the small open circles on the model tracks denote
    metallicities in \Oabundance. The dotted lines connect the 
    photoionization model curves with the same metallicity value.
  }
  \label{fig:o3o2r23_solo}
\end{figure*}

\section{Testing FMR with Our SDSS Sample}
\label{sec:testing_sample}

Before we present our analysis results, we test whether our SDSS sample 
reproduces the FMR \citep{mannucci2010}.
For this purpose, we derive metallicities based on the 
\citeauthor{maiolino2008}'s (\citeyear{maiolino2008}) 
empirical relations, as adopted by \citet{mannucci2010}
(Section \ref{sssec:sample_z0_sdss}).
It should be noted that our main discussion is based 
on metallicities derived with ionization parameter 
from the photoionization models of \citet{KD2002} 
(e.g., Section \ref{ssec:interpretation_o3o2r23}).
In Figure \ref{fig:fmr_Man10}, three panels from left to right show the 
$M_{\star}$-metallicity, SFR-metallicity, and $\mu_{0.32}$-metallicity
plots with our SDSS sample. The variable $\mu_{\alpha}$ is defined as a 
combination of $M_{\star}$ and SFR,
\begin{equation}
  \mu_{\alpha} = \log(M_{\star})-\alpha\log({\rm SFR}),
  \label{eq:mu_alpha}
\end{equation}
where $\alpha$ is a free parameter.
\citet{mannucci2010} obtain $\alpha=0.32$ that minimizes the scatter of 
their SDSS galaxies in the $\mu_{\alpha}$-metallicity plane.
The colored lines in each panel represent a mean metallicity of our 
SDSS galaxies for subsamples defined with SFR (left and right panels) 
and $M_{\star}$ (center panel) in a $0.15$\,dex range. We display 
subsamples containing $>50$ galaxies.

In the right panel of Figure \ref{fig:fmr_Man10}, the solid black 
curve indicates the quadratic best-fit of our SDSS sample, while the
dotted curve presents the original FMR 
\citep{mannucci2010,mannucci2011}. The original FMR is expressed by a 
combination of a linear and a 4th-order polynomial functions 
\citep{mannucci2011}.
From this plot, the FMR of our SDSS sample is almost identical to that
of \citet{mannucci2010,mannucci2011}. Therefore, we conclude that our 
SDSS sample successfully reproduces the original FMR of 
\citet{mannucci2010}%
\footnote{
In our SDSS sample, we find a value of $\alpha=0.30$ that minimizes the 
scatter in the $\mu_{\alpha}$-metallicity plane. However, the difference 
of residual dispersion of metallicity between $\alpha=0.32$ and $0.30$ 
is negligibly small, $0.0251$ and $0.0255$\,dex, respectively.
}.

\section{Ionization State of Inter-Stellar Medium in Galaxies}
\label{sec:o3o2r23}

\subsection{\OIII$/$\OII\ vs. $R23$-index Diagram}
\label{ssec:o3o2r23_solo}

\citet{nakajima2013} have demonstrated that \OIII$/$\OII\ vs. $R23$-index 
diagram is useful to investigate ionization parameter and metallicity
of galaxies (see also \citealt{KD2002}).
Figure \ref{fig:o3o2r23_solo} is the diagram of \OIII$/$\OII\ ratio and 
$R23$-index with our local, intermediate-$z$ ($z\sim 1$), and high-$z$ 
($z\sim 2-3$) galaxies.
The underlying gray dots are the SDSS galaxies. 
Figure \ref{fig:o3o2r23_solo} indicates that galaxies with a higher 
$R23$-index have a higher \OIII$/$\OII\ ratio.
The red and blue diamonds denote $z\sim 2-3$ LAEs and LBGs, respectively.
Compared to the SDSS galaxies, these high-$z$ galaxies clearly present 
an \OIII$/$\OII\ ratio higher than most of galaxies by a factor of 
$\ga 10$. The majority of $z\sim 1$ galaxies (the gray triangles) are 
placed at the middle between the SDSS galaxies and the $z\sim 2-3$ LBGs.
LBGs at $z\ga 3$ (blue diamonds with a circle) have an \OIII$/$\OII\ 
ratio higher than those at $z\la 3$ on average.
In Figure \ref{fig:o3o2r23_solo}, LAEs have the highest \OIII$/$\OII\ 
ratio. This trend is first claimed by \citet{nakajima2013}, but we 
confirm this trend with our LAE and LBG samples significantly larger 
than those used in \citet{nakajima2013}.
The two local galaxies of GPs (green open circles) and LBAs 
(purple circles) are placed near the high-$z$ galaxies. These galaxies 
present the highest $R23$-index and \OIII$/$\OII\ ratio among the local 
galaxies. Interestingly, the LyC leakers (orange pentagons) have values 
of \OIII$/$\OII\ ratio and $R23$-index similar to those of GPs, LBAs, 
and high-$z$ galaxies.
It should be noted that some LAEs have a remarkably high \OIII$/$\OII\ 
ratio that cannot be found in the local galaxies.

\subsection{Physical Origin of Large \OIII$/$\OII\ ratio}
\label{ssec:origin_o3o2}

We investigate the physical origin of large \OIII$/$\OII\ ratios found in 
Section \ref{ssec:o3o2r23_solo}, using photoionization models.
There are five physical parameters of nebulae that change the \OIII$/$\OII\ 
ratio,
a) spectral shape of ionizing source,
b) gas temperature,
c) metallicity,
d) ionization parameter, and
e) gas density.
First, if the incident ionizing source includes an AGN whose spectrum is 
hard, high-ionization lines are stronger than low-ionization lines.
However, since our samples are free from a significant AGN contamination
(Section \ref{sssec:sample_z0_sdss}), the ionizing source difference does 
not significantly change the \OIII$/$\OII\ ratio.
Second, gas temperature changes the emissivity of \OIII\ and \OII\ 
lines. Gas temperature is determined by the metal abundance, because 
coolings are dominated by collisionally excited metal ions such as 
O$^{+}$, O$^{2+}$, N$^{+}$ etc. We simply regard metallicity as a 
control parameter of temperature, and investigate the temperature 
dependence on the basis of gas metallicity. 
Thus, there are three key parameters, metallicity, ionization 
parameter, and gas density, that affect the \OIII$/$\OII\ ratio.

In Figure \ref{fig:o3o2r23_solo}, we plot the photoionization models
given by \citet{KD2002} with the black curves. The models predict
line ratios at a given metallicity and ionization parameter
with a fixed gas pressure%
\footnote{
For electron temperatures of $10^4$\,K. 
This pressure corresponds to a density of order $10$\,cm$^{-3}$.
}
of $P/k=10^5$\,cm$^{-3}$\,K 
assuming isobaric conditions.
The models in Figure \ref{fig:o3o2r23_solo} indicate that the
\OIII$/$\OII\ ratio and $R23$-index depend on metallicity and 
ionization parameter.

The models of Figure \ref{fig:o3o2r23_solo} present three important 
trends on the \OIII$/$\OII\ vs. $R23$-index diagram;
\begin{enumerate}
  \renewcommand{\theenumi}{(\arabic{enumi})}
  \item \OIII$/$\OII\ ratio increases with ionization parameter.
    \label{item:o3o2_qion}
  \item \OIII$/$\OII\ ratio decreases with metallicity.
    \label{item:o3o2_Z}
  \item $R23$-index increases in a low-metallicity regime 
    (low-$Z$ branch) from \Oabundance\ $=7.5$ to $\sim 8.5$,
    and decreases in a high-metallicity regime (high-$Z$ branch)
    from \Oabundance\ $\sim 8.5$ towards a higher metallicity.
    \label{item:r23_Z}
\end{enumerate}
The trend of \ref{item:o3o2_qion} is explained by the boost of 
high-energy ionizing photons as the ionization parameter increases. 
Such high-energy photons produce more abundant \OIII.
The trend of \ref{item:o3o2_Z} is caused by a decrease of gas
temperature in an \HII-region. As gas metallicity increases,
gas temperature drops due to cooling of metal lines.
For gas clouds with high metallicity, the far-infrared
fine-structure cooling of O$^{2+}$ is more efficient
than those in optical wavelength (e.g., \citealt{CL2001,dopita2000}).
Thus, in optical wavelength, \OIII\ emission becomes faint,
while \OII\ emission does not change dramatically, which makes 
a decrease of \OIII$/$\OII\ ratio to a higher metallicity.
Similarly, the trend of \ref{item:r23_Z} in the high-$Z$ branch of 
\Oabundance\ $\ga 8.5$ is explained by the weak optical \OIII$+$\OII\ 
emission with more dominant far-infrared fine-structure cooling.
Thus, the $R23$-index decreases with metallicity. On the other hand, 
in the low-$Z$ branch of \Oabundance\ $\la 8.5$, optical collisional 
excitation lines including \OIII\ and \OII\ are the most efficient 
cooling radiation, $R23$-index increases as oxygen abundance, i.e., 
metallicity, increases.
To summarize, the \OIII$/$\OII\ ratio depends strongly on ionization 
parameter and moderately on metallicity. $R23$-index is mainly 
governed by metallicity, but also depends, albeit weakly, on 
ionization parameter.

In addition to metallicity and ionization parameter, emission line 
ratios would be changed by gas density.
In fact, the recent study of \citet{shirazi2013} argues that  
a high gas density would be an origin of the large \OIII$/$\OII\ 
ratio of high-$z$ galaxies. 
This argument is based on the following ideas.
The ionization parameter is defined by
\begin{equation}
  q_{\rm ion} = \frac{Q_{{\rm H}^{0}}}{4\pi R_{\rm s}^{2} n_{\rm H}},
  \label{eq:qion}
\end{equation}
where $Q_{{\rm H}^{0}}$ is the number of hydrogen ionizing photons 
per second, $R_{\rm s}$ is the \Stromgren\ radius, and $n_{\rm H}$ 
is the total hydrogen density.
The hydrogen ionizing photon production rate of $Q_{{\rm H}^{0}}$
is balanced by a recombination rate
under the assumption of ionization equilibrium,
\begin{equation}
  Q_{{\rm H}^{0}} = \frac{4}{3}\pi R_{\rm s}^{3} n_{\rm H}^{2} \alpha_{B} \epsilon,
  \label{eq:equilibrium}
\end{equation}
where $\alpha_{B}$ is a coefficient of the total hydrogen
recombination to the $n>1$ levels, and $\epsilon$ is the volume filling 
factor of the ionized gas.
The combination of Equations (\ref{eq:qion}) and (\ref{eq:equilibrium})
yields
\begin{equation}
  q_{\rm ion}^{3} \propto Q_{{\rm H}^{0}}\, n_{\rm H}\, \epsilon^{2}
  \label{eq:qion3_n1}
\end{equation}
(see also \citealt{CL2001,shirazi2013}).
By the comparison with low and high redshift galaxies with similar 
$M_{\star}$ and specific SFR
(and thus with similar metallicity under the assumption of the FMR 
being in place), 
\citet{shirazi2013} have found that the high-$z$ 
galaxies show an \OIII$/$\OII\ ratio higher than low-$z$ counterparts by
$\sim 0.5$\,dex, indicating that high-$z$ galaxies have a higher \qion\ 
than the low-$z$ counterparts by $\sim 0.3$\,dex.
Based on Equation (\ref{eq:qion3_n1}), \citet{shirazi2013} have argued 
that $n_{\rm H}$ increases by a factor of $\sim 8$ to high-$z$, assuming 
that high-$z$ galaxies and their low-$z$ counterparts have similar 
$Q_{{\rm H}^{0}}$ and $\epsilon$ values.
Indeed, \citet{shirazi2013} have confirmed that high-$z$ galaxies 
tend to have a high electron density, which is measured directly 
from emission lines.

Because ionization parameter is determined by 
$Q_{{\rm H}^{0}}$, $n_{\rm H}$, and $\epsilon$
(Equation \ref{eq:qion3_n1}), ionization parameter is a fundamental 
parameter of photoionization models, which allows the 
potential evolutions of $n_{\rm H}$, $Q_{{\rm H}^{0}}$, and $\epsilon$.
The assumption of no evolution of IMF would also be an open 
question (e.g., \citealt{dave2008,ND2013}).
Therefore, we interpret the \OIII$/$\OII\ ratio as a function
of ionization parameter and metallicity in this paper.
Hereafter, we study the 
relation between \OIII$/$\OII\ ratio and $R23$-index with various 
ionization parameters and metallicities in the following analysis.

\subsection{Interpretation of the \OIII$/$\OII\ and $R23$-index Relation}
\label{ssec:interpretation_o3o2r23}

In this section, we evaluate the average values of our galaxies' 
ionization parameter and metallicity that are sensitive to \OIII$/$\OII\ 
ratio and $R23$-index measurements, using the method given by 
\citet{KK2004}.
\citet{KK2004} use the photoionization model of \citet{KD2002}, showing 
an equation of ionization parameter,
\begin{eqnarray}
   && \log(q_{\rm ion})=\{ 32.81-1.153y^2 \nonumber\\
   && \ \ \ +[12+\log({\rm O/H})](-3.396-0.025y+0.1444y^2) \} \nonumber\\
   && \ \ \ \times \{ 4.603-0.3119y-0.163y^2 \nonumber\\
   && \ \ \ +[12+\log({\rm O/H})](-0.48+0.0271y+0.02037y^2) \}^{-1}, \nonumber\\
   &&
  \label{eq:qion_KK04}
\end{eqnarray}
where $y=\log([{\rm O}\,{\scriptstyle{\rm III}}]\lambda\lambda 5007,4959/[{\rm O}\,{\scriptstyle{\rm II}}]\lambda 3727)$, and equations of metallicity,
\begin{eqnarray}
  && 12+\log({\rm O/H})_{\rm low} = 9.40+4.65x-3.17x^2 \nonumber\\
  && \ \ \ \ \ \ \ \ \ \ \ \ -\log(q_{\rm ion})\times (0.272+0.547x-0.513x^2),
  \label{eq:Z_l_KK04} \\
  && 12+\log({\rm O/H})_{\rm high} = 9.72-0.777x-0.951x^2 \nonumber\\
  && \ \ \ \ \ \ \ \ \ \ \ \ -0.072x^3-0.811x^4 \nonumber\\
  && \ \ \ \ \ \ \ \ \ \ \ \ -\log(q_{\rm ion})\times (0.0737-0.0713x-0.141x^2 \nonumber\\
  && \ \ \ \ \ \ \ \ \ \ \ \ +0.0373x^3-0.058x^4),
  \label{eq:Z_h_KK04}
\end{eqnarray}
where $x=\log(R23)$ and the subscript low (high) corresponds to
a metallicity value in the low-$Z$ (high-$Z$) branch of metallicity.
We use dust-corrected \OIII$/$\OII\ ratio and $R23$-index.
Note that, without other metallicity indicators, this method does not 
provide a single solution, but multiple solutions from low-$Z$ and 
high-$Z$ branches. The demarcation metallicity between the two branches
is \Oabundance\ $=8.4$.
We use Equations (\ref{eq:qion_KK04})--(\ref{eq:Z_h_KK04}) to estimate
a metallicity and ionization parameter iteratively until the metallicity
converges.
We evaluate errors of metallicity and ionization parameter by Monte Carlo 
simulations. For each galaxy sample, we obtain the average values of 
$R23$-index, \OIII$/$\OII\ ratio, and their uncertainties. We assume that 
for each of the flux ratios, the probability distribution of the true 
value is a Gaussian with its sigma equal to the observed uncertainty in 
the flux ratio. We randomly select a set of $R23$-index and \OIII$/$\OII\ 
ratio following these Gaussian probability distributions, and calculate 
metallicity and ionization parameter with the set of $R23$-index and 
\OIII$/$\OII\ ratio based on Equations 
(\ref{eq:qion_KK04})--(\ref{eq:Z_h_KK04}). We repeat the process $500$ 
times, and define one sigma errors of metallicity and ionization parameter 
by the $68$ percentile of the distribution.

\begin{figure}
  \centerline{
    \includegraphics[width=0.95\columnwidth]{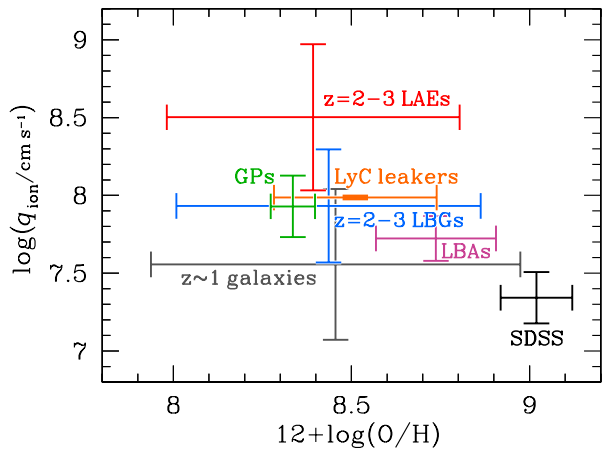}
  }
  \caption{
    Average ionization parameter and metallicity for our galaxy
    samples (Table \ref{tbl:Z_qion}).
  }
  \label{fig:Z_qion}
\end{figure}

\begin{table}
  \centering
  \caption{Average ionization parameter, $\log($\qion$)$, and
    metallicity, \Oabundance, ranges for our galaxy samples estimated 
    from the \OIII$/$\OII\ ratio and $R23$-index.
    }
  \label{tbl:Z_qion}
  \begin{tabular}{@{}lcc@{}}
    \hline
    Sample &
    $\log($\qion$)$ &
    \Oabundance\ \\
    \hline
    SDSS &
    $7.18$--$7.51$ &
    $8.92$--$9.12$ \\
    $z\sim 1$ galaxies &
    $7.07$--$8.04$ &
    $7.94$--$8.97$ \\
    LyC leakers &
    $7.97$--$8.00$ &
    $8.28$--$8.74$ \\
    LBAs &
    $7.58$--$7.87$ &
    $8.57$--$8.91$ \\
    GPs &
    $7.73$--$8.13$ &
    $8.27$--$8.40$ \\
    $z=2-3$ LBGs &
    $7.57$--$8.29$ &
    $8.01$--$8.87$ \\
    $z=2-3$ LAEs &
    $8.04$--$8.97$ &
    $7.98$--$8.81$ \\
    \hline
  \end{tabular}
\end{table}

The estimated values are summarized in Table \ref{tbl:Z_qion} and shown 
in Figure \ref{fig:Z_qion}. 
We list the ranges of values in both the high-$Z$ and low-$Z$ branches
for the intermediate and high-$z$ galaxy samples but for the local galaxy
samples.
For the SDSS sample, only the values in the high-$Z$ branch are given, 
since the average metallicity for the low-$Z$ branch,
\Oabundance\ $=7.7-8.1$, is significantly lower than that estimated with 
$R23$-index and $N2$-index (Figure \ref{fig:fmr_Man10}). We have checked 
that the SDSS high-$Z$ branch metallicity of \Oabundance\ $=8.92-9.12$ 
(Table \ref{tbl:Z_qion}) is consistent with our estimate in Section 
\ref{sec:testing_sample}, as well as those estimated in previous studies
(e.g., \citealt{tremonti2004,mannucci2010}).
For GPs, LBAs, and LyC leakers, we calculate the metallicities and 
ionization parameters for individual objects, whose two-branch solutions
are distinguished by other metallicity indicators of 
gas temperature and/or $N2$-index 
(see also Section \ref{ssec:fmr_z3}). 
Their $68$ percentile of the distributions are listed 
in Table \ref{tbl:Z_qion} and shown in Figure \ref{fig:Z_qion}.

Although only weak constraints are placed on metallicity due to
the degeneracy of the two branch solutions, the difference of 
ionization parameters between galaxy samples is clearly found in 
Figure \ref{fig:Z_qion}. 
The SDSS galaxies have the average value of 
$\log($\qion$/$\cms$) \sim 7.3$ (see also e.g., \citealt{dopita2006a}).
In contrast, high-$z$ galaxies have an ionization parameter of 
$\log($\qion$/$\cms$) \sim 7.6-9.0$. 
High-$z$ LBGs have an ionization parameter higher than the SDSS 
galaxies by a factor of $\sim 4$, and high-$z$ LAEs show the highest 
ionization parameter among the galaxy samples, which is about an order 
of magnitude higher than the SDSS galaxies. 
In the local galaxies, GPs and the LyC leakers exhibit 
ionization parameters much higher than the SDSS galaxies. Similarly, 
the ionization parameter of LBAs is high. These extreme local 
populations of GPs, LyC leakers, and LBAs have an ionization parameter 
as high as LBGs and LAEs at high redshifts. Moreover, the ionization 
parameter of GPs is comparable to that of LAEs. In this sense, GPs 
could be local counterparts of high-$z$ LAEs.

\begin{figure}
  \centerline{
    \includegraphics[width=0.87\columnwidth]{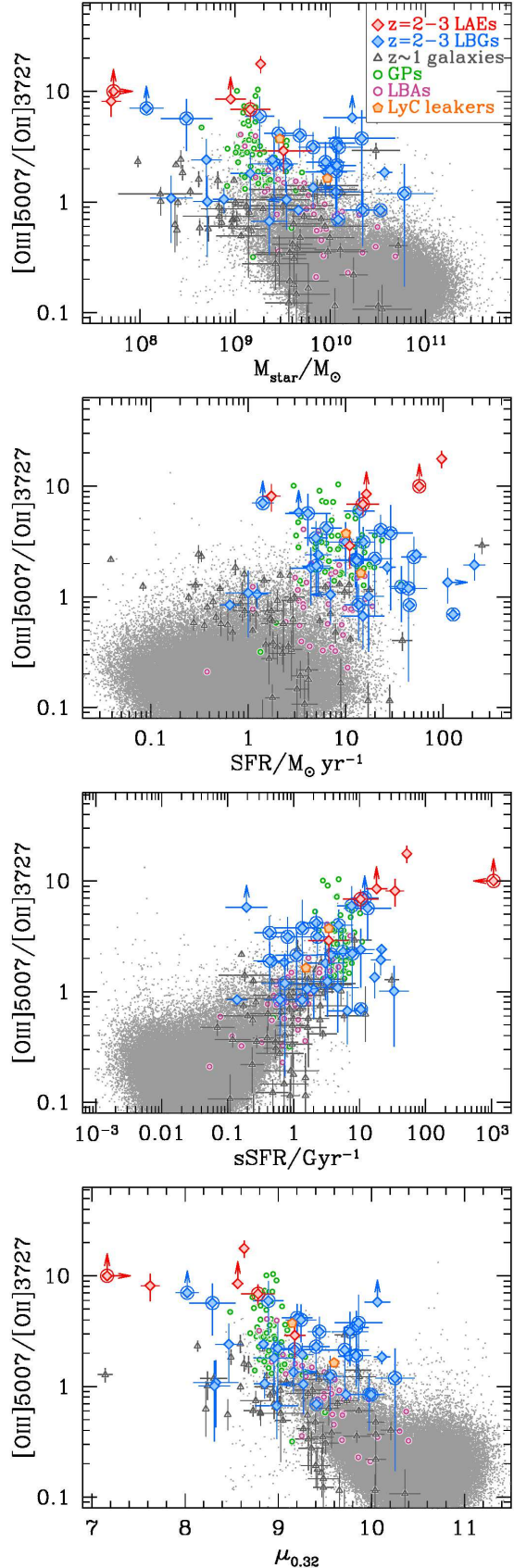}
  }
  \caption{
    From top to bottom, dependence of \OIII$/$\OII\ on $M_{\star}$, SFR,
    sSFR, and $\mu_{0.32}$.
    The symbols and colors are the same as those used in Figure
    \ref{fig:o3o2r23_solo}.
  }
  \label{fig:o3o2_properties}
\end{figure}

Figure \ref{fig:o3o2r23_solo} suggests that there is a variation of 
ionization parameter in the SDSS sample.
In fact, the SDSS galaxy 
distribution sequence departs from the model curve of 
$\sim 2\times 10^7$\,\cms.
In the large $R23$-index regime, we recognize the slope of the SDSS 
galaxy sequence steeper than that of the photoionization model 
predictions. This trend indicates that galaxies with a low metallicity 
of \Oabundance\ $\sim 8.5$ have a high ionization parameter 
(e.g., \citealt{dopita2006a,nagao2006}).
If the $M_{\star}$--$Z$ relation is in place 
(e.g., \citealt{tremonti2004}), this would imply that low-$M_{\star}$ 
galaxies have a high ionization parameter.
We examine the possible dependencies of $M_{\star}$ and other galaxy
global properties on ionization parameters in Sections 
\ref{ssec:o3o2_properties} and \ref{ssec:o3o2r23_FIR}.
Similarly, the important characteristics can be found in the small 
$R23$-index regime. The sequence of SDSS galaxies presents an upturn 
from high to low $R23$-index values. This trend is probably caused by 
the presence of a hidden AGN activity in a galaxy. In this regime, 
the SDSS galaxies departing from the model curve would be also 
dominated by objects with a low SFR and a high $M_{\star}$
(Section \ref{ssec:o3o2r23_FIR}; see also \citealt{dopita2006a}).

\begin{figure*}
  \centerline{
    \includegraphics[width=0.84\textwidth]{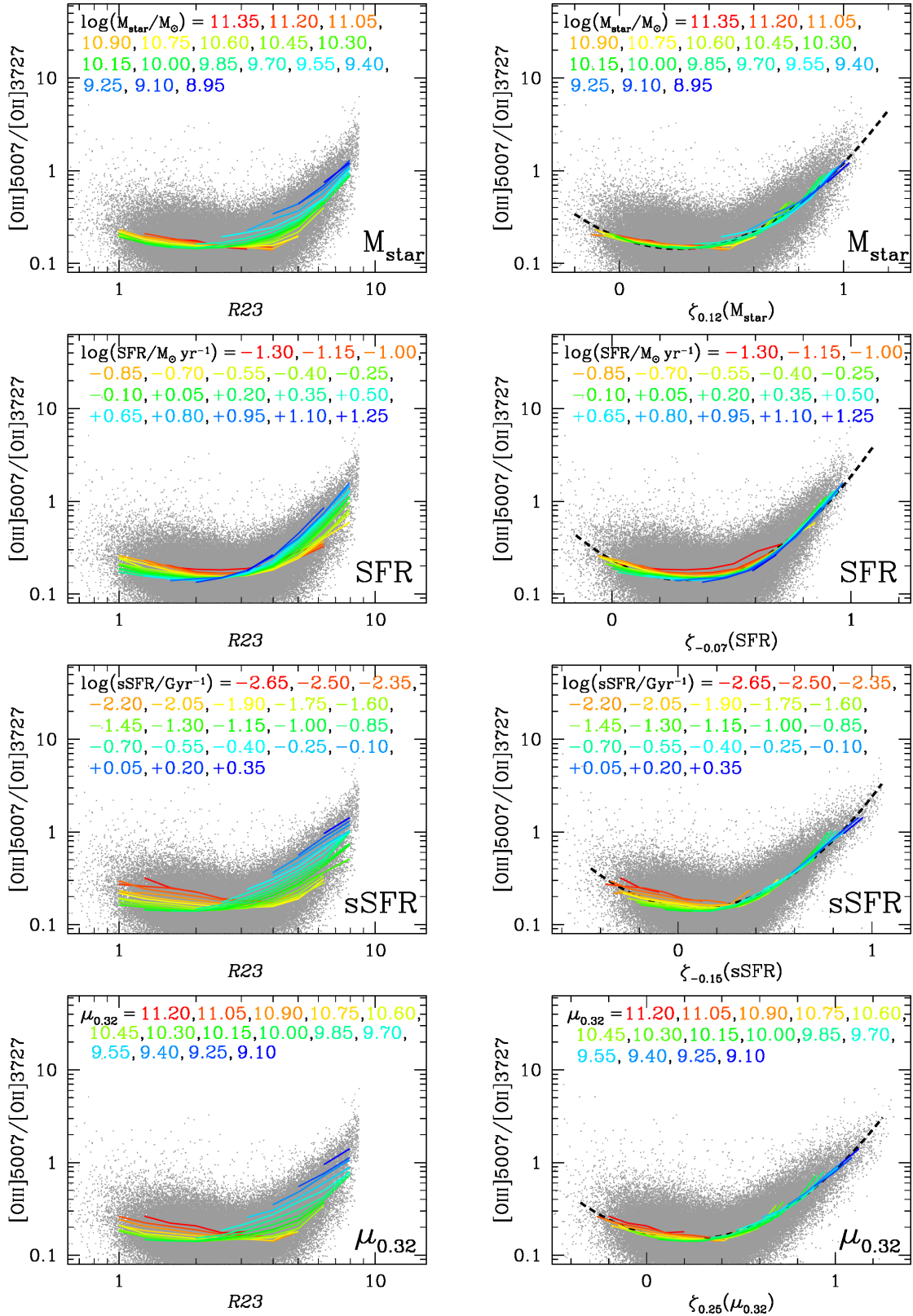}
  }
  \caption{
    {\em Left:} \OIII$/$\OII\ vs. $R23$-index diagrams for the SDSS 
    galaxies. The gray dots denote all of the SDSS galaxies.
    From top to bottom, the colored lines represent the median 
    \OIII$/$\OII\ values for subsamples defined by a 3rd quantity of 
    $M_{\star}$, SFR, sSFR, and $\mu_{0.32}$.
    The definition of the subsamples are indicated at the top of 
    each panel. Subsamples including $>50$ galaxies are presented.
    {\em Right:} Same as the left panels, but for \OIII$/$\OII\ vs. 
    $\zeta_{\beta}\,(\Theta)$ diagrams. The black dashed-line shows the
    quadratic best-fit (Equation (\ref{eq:bestfit_o3o2r23_3rdparam})
    with coefficients given in Table \ref{tbl:coefficients_zeta}).
  }
  \label{fig:o3o2r23_all}
\end{figure*}

\subsection{Dependence of \OIII$/$\OII\ on $M_{\star}$ and SFR}
\label{ssec:o3o2_properties}

In Section \ref{ssec:interpretation_o3o2r23}, we have estimated the 
typical ionization parameters for our galaxy samples from the comparisons 
of the photoionization models with the measurements of \OIII$/$\OII\ and 
$R23$-index, and compared them (Figure \ref{fig:Z_qion}). 
However, these comparisons do not focus on possible dependencies on 
galaxy global properties such as $M_{\star}$ and SFR.
Figure \ref{fig:o3o2r23_solo} presents a relatively large scatter in 
\OIII$/$\OII\ ratios at a given $R23$-index in one galaxy sample.
This scatter could be originated from the galaxy global properties.
To address the issue, we examine the dependencies of the \OIII$/$\OII\ 
ratio on the galaxy global properties.

Figure \ref{fig:o3o2_properties} shows \OIII$/$\OII\ ratios as a function 
of $M_{\star}$, SFR, specific SFR (sSFR; SFR divided by $M_{\star}$), and 
$\mu_{0.32}$ (Equation (\ref{eq:mu_alpha}) with $\alpha=0.32$).
Figure \ref{fig:o3o2_properties} indicates that the \OIII$/$\OII\ ratio 
correlates with the galaxy global properties of $M_{\star}$, SFR, sSFR, 
and $\mu_{0.32}$. A high \OIII$/$\OII\ ratio is found in low $M_{\star}$, 
low $\mu_{0.32}$, and high sSFR galaxies.
A similar trend is initially proposed by \citet{brinchmann2008}, 
but we find these trends in local extreme populations and high-$z$
galaxies.
There is an anti-correlation between \OIII$/$\OII\ and metallicity 
(Section \ref{ssec:origin_o3o2}). Thus, the dependence of \OIII$/$\OII\ 
ratio on $M_{\star}$ and $\mu_{0.32}$ is reasonable, because
the $M_{\star}$--$Z$ relation and the FMR are in place.
In this way, the \OIII$/$\OII\ ratio is tightly related to the galaxy 
global properties, while this ratio is an indicator of ionization 
parameter and metallicity. In the next section, we investigate 
dependencies of the galaxy global properties on ionization parameter 
and metallicity with the SDSS sample on the \OIII$/$\OII\ vs. 
$R23$-index diagram.

Before moving to the next section, we comment here on the variations
of global properties for the galaxies shown in Figure
\ref{fig:o3o2_properties}.
First, GPs are the least massive and the most actively star-forming 
galaxies in the local universe. Their low $\mu_{0.32}$ values are 
probably due to their low metallicities (e.g., \citealt{amorin2010}). 
LBAs are the next extreme population. 
The LyC leakers have SFRs as high as GPs and LBAs.
High-$z$ galaxies have a $M_{\star}$ comparable with the SDSS galaxies, 
but a SFR higher than these local galaxies. 
This tendency is consistent with the evolution of the 
star-formation main sequence (e.g., \citealt{daddi2007}).
The SFRs and $M_{\star}$ of 
high-$z$ galaxies are little higher than those of GPs and LBAs, while 
the sSFR of high-$z$ galaxies are comparable with those of GPs and LBAs. 
LAEs exhibit remarkable properties on average, having the least 
$M_{\star}$ and the highest SFR among galaxies at any redshifts.

\subsection{Fundamental Ionization Relation}
\label{ssec:o3o2r23_FIR}

As shown in Figure \ref{fig:o3o2r23_solo}, 
the \OIII$/$\OII\ vs. $R23$-index diagram allows us to study ionization 
state of hot ISM in galaxies.
In Section \ref{ssec:o3o2_properties}, we discuss that the \OIII$/$\OII\ 
ratio depends on four galaxy global properties of $M_{\star}$, SFR, sSFR, 
and $\mu_{0.32}$. The large scatter found in Figure \ref{fig:o3o2r23_solo}
would be explained by the differences of the galaxy global properties.

In order to investigate the dependencies of $M_{\star}$, SFR, sSFR,
and $\mu_{0.32}$ on the \OIII$/$\OII\ vs. $R23$-index plane, we make 
subsamples of our SDSS sample. We choose one quantity from the four 
galaxy global properties. The chosen quantity is referred to as a 
{\lq}3rd quantity{\rq} hereafter. A bin size of a subsample is $0.15$\,dex 
in the 3rd quantity. For a given subsample, we calculate a median 
\OIII$/$\OII\ ratio of galaxies in a range of $\Delta$$\log$$R23=0.1$.
We use subsamples with $>50$ galaxies for our analysis.
The left panels of Figure \ref{fig:o3o2r23_all} show the subsamples of 
$M_{\star}$, SFR, sSFR, and $\mu_{0.32}$ on the \OIII$/$\OII\ vs. 
$R23$-index plane.
Figure \ref{fig:o3o2r23_all} indicates that the relation between
\OIII$/$\OII\ and $R23$-index depends on the galaxy global properties.
A high \OIII$/$\OII\ ratio can be found in less massive, more 
efficiently star-forming, and probably more metal-poor galaxies
\citep{dopita2006a}.
We introduce a parameter, $\zeta_{\beta}\,(\Theta)$, which combines
$R23$-index and one of the 3rd quantities:
\begin{equation}
  \zeta_{\beta}\,(\Theta) = \log(R23)-\beta \times \Theta,
  \label{eq:zeta}
\end{equation}
where
$\beta$ is a free parameter and
$\Theta=\log(M_{\star}/10^{10}\,M_{\odot})$,
$\log({\rm SFR}/M_{\odot}\,{\rm yr}^{-1})$,
$\log({\rm sSFR}/{\rm Gyr}^{-1})$, and
$(\mu_{0.32}-10.0)$
for a 3rd quantity of $M_{\star}$, SFR, sSFR, and $\mu_{0.32}$,
respectively.
We calculate a scatter of the SDSS subsamples in the relation of 
\OIII$/$\OII\ and $\zeta_{\beta}\,(\Theta)$ by varying $\beta$
for each of the 3rd quantities, $\Theta$.
As a result, the scatter is minimized
at $\beta=(0.12,\, -0.07,\, -0.15,\, 0.25)$ for
$\Theta$ of ($M_{\star}$, SFR, sSFR, $\mu_{0.32}$).
The right panels of Figure \ref{fig:o3o2r23_all} present our SDSS 
galaxies on the \OIII$/$\OII\ and $\zeta_{\beta}\,(\Theta)$ plane.
The median values of the SDSS subsamples are fitted by a quadrature:
\begin{equation}
  \log([{\rm O}\,{\scriptstyle{\rm III}}]/[{\rm O}\,{\scriptstyle{\rm II}}])
   = a_{0}+a_{1}\zeta_{\beta}\,(\Theta)+a_{2}\zeta_{\beta}^{2}\,(\Theta),
  \label{eq:bestfit_o3o2r23_3rdparam}
\end{equation}
where $a_{0}$--$a_{2}$ are free parameters.
The best-fit parameters are given in Table \ref{tbl:coefficients_zeta}.
We show the best-fit quadratures in the right panels of 
Figure \ref{fig:o3o2r23_all}. These plots suggest that the 3rd 
quantity can reduce the systematic dispersions.
However, it is unclear whether the new parameter of $\beta$ reduces 
the dispersion by a physical reason or a simple statistical reason.
Here we use $\mu_{\alpha}$ defined in Equation (\ref{eq:mu_alpha}) as 
a 3rd quantity and search for an optimal combination of $M_{\star}$
and SFR by varying parameters of both $\alpha$ and $\beta$
\footnote{
Although $\mu_{\alpha}$ does not represent a function of SFR with no 
dependence of $M_{\star}$, we have found in our analysis that a $\zeta$
function with $\Theta = \Theta({\rm SFR})$ provides the best-fit value 
of $\beta\simeq 0$ (Table \ref{tbl:coefficients_zeta}), which indicates 
that a SFR alone does not reduce the scatter of the SDSS galaxy 
distribution.
}.
Note that the parameter $\alpha$ denotes the best combination of 
$M_{\star}$ and SFR for metallicity \citep{mannucci2010}, while the 
parameter $\beta$ defines the best combination of these quantities for 
both metallicity and ionization parameter.
Based on Equation (\ref{eq:zeta}) with $\Theta=(\mu_{\alpha}-10.0)$,
we calculate $\chi^2$ of the free parameters, $\alpha$ and $\beta$.
Here we assume that $1\sigma$ \OIII$/$\OII\ dispersions of our 
subsamples are originated from measurement errors. These $1\sigma$ 
dispersions are used to obtain the $\chi^2$ values.

\begin{figure}
  \centerline{
    \includegraphics[width=0.95\columnwidth]{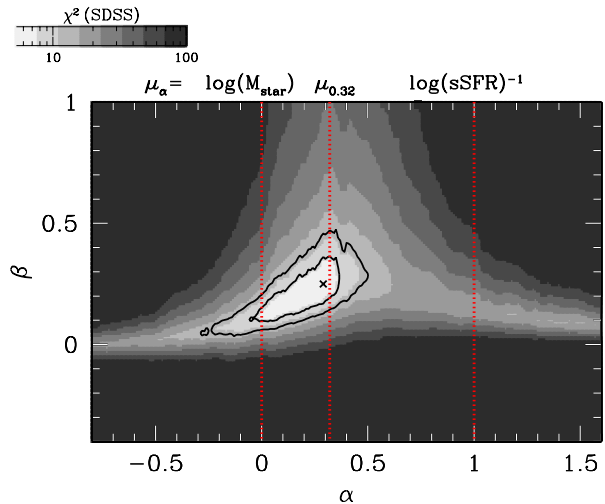}
  }
  \caption{
    $\chi^2$ contours of free parameters, $\alpha$ and $\beta$,
    that are used in the \OIII$/$\OII-$\zeta_{\beta}\,(\mu_{\alpha})$
    relation of our SDSS sample. The gray scale represents a $\chi^2$ 
    value as shown in the gray bar at the top. A darker color indicates 
    a larger $\chi^2$ value. The black solid contours represent the 
    $1\sigma$ (inner) and $2\sigma$ (outer) confidence levels, and the 
    cross denotes the minimum $\chi^2$ of $5.3$ at 
    ($\alpha$, $\beta$)=($0.29$, $0.25$). The values of $\alpha=0$ and $1$ 
    correspond to $\mu_{\alpha}=\log(M_{\star})$ and $\log({\rm sSFR})^{-1}$,
    respectively.
  }
  \label{fig:alpha_beta}
\end{figure}

\begin{table}
  \centering
  \caption{Best-fit values for the quadrature of Equation 
    (\ref{eq:bestfit_o3o2r23_3rdparam}).}
  \label{tbl:coefficients_zeta}
  \begin{tabular}{@{}lcccc@{}}
    \hline
    3rd quantity &
    $\beta$ &
    $a_{0}$ &
    $a_{1}$ &
    $a_{2}$ \\
    \hline
    $M_{\star}$ &
    $+0.12$ &
    $-0.722$ &
    $-0.923$ &
    $+1.726$ \\
    SFR &
    $-0.07$ &
    $-0.634$ &
    $-1.427$ &
    $+2.331$ \\
    sSFR &
    $-0.15$ &
    $-0.810$ &
    $-0.267$ &
    $+1.458$ \\
    $\mu_{0.32}$ &
    $+0.25$ &
    $-0.772$ &
    $-0.536$ &
    $+1.232$ \\
    \hline
  \end{tabular}
\end{table}

\begin{figure*}
  \centerline{
    \includegraphics[width=0.75\textwidth]{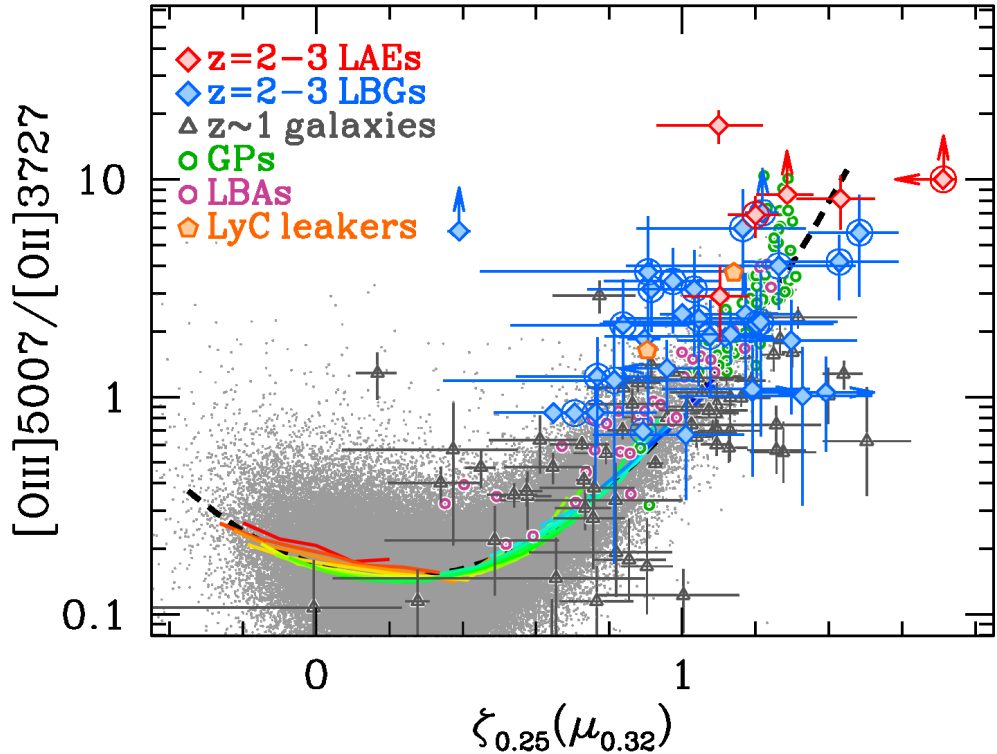}
  }
  \caption{
    \OIII$/$\OII\ vs. $\zeta_{0.25}\,(\mu_{0.32})$ relation (FIR). The
    color code for the lines are the same as those at the
    bottom right panel of Figure \ref{fig:o3o2r23_all}.
    In this plot, we compare the relation of the SDSS galaxies
    with all of our galaxy samples, high-$z$
    galaxies, GPs, LBA, and the LyC leakers.
  }
  \label{fig:o3o2r23_mu_f_solo}
\end{figure*}

Figure \ref{fig:alpha_beta} shows $\chi^2$ values of $\alpha$ and 
$\beta$.
The minimum $\chi^2$ is provided by the parameters of
$(\alpha,\, \beta)=(0.29,\, 0.25)$.
The best-fit parameters are $\alpha=0.29^{+0.04}_{-0.26}$ and
$\beta=0.25^{+0.06}_{-0.12}$.
Interestingly, the best-fit value of $\alpha=0.29^{+0.04}_{-0.26}$ 
is consistent with the one of the FMR ($\alpha=0.32$).
Thus, $\mu_{0.32}$ ($\mu_\alpha$ for $\alpha=0.32$) is the best 3rd 
quantity to describe the relation.
Moreover, we have newly constrained the $\beta$ parameter that is 
tightly related to ionization parameter. The best-fit $\beta$ rules 
out $\beta=0$ at the $>95$\,\% confidence level. The results of our 
analysis suggest that ionization parameters given with the 
\OIII$/$\OII\ vs. $R23$-index diagram is important to characterize 
galaxies, and that ionization parameters are closely related to the 
galaxy global properties of $M_{\star}$ and SFR as well as 
metallicity.

We compare our various galaxy samples with the best-fit function of 
the SDSS galaxies in Figure \ref{fig:o3o2r23_mu_f_solo}. This figure 
is the same as the bottom right panel of Figure \ref{fig:o3o2r23_all}, 
but with high-$z$ galaxies and local extreme populations, GPs, LBAs, 
and the LyC leakers. Although the best-fit function is derived with the 
SDSS sample alone, high-$z$ galaxies and local extreme populations 
follow the same relation or fall on the extrapolation of the function.
It should be noted that no significant differences are found between 
galaxies at $z\sim 2$ and $3$.
The relation of best-fit function between \OIII$/$\OII\ and 
$\zeta_{\beta}\,(\Theta)$ (Figure \ref{fig:o3o2r23_mu_f_solo}) is thus 
thought to be a fundamental relation consisting of ionization parameter, 
metallicity, $M_{\star}$, and SFR over cosmic time. Hereafter we refer to 
the relation as fundamental ionization relation (FIR).

\section{Discussion}
\label{sec:discussion}

\subsection{Evolution of Ionization State}
\label{ssec:qion_evolution}

We have confirmed the trend that high-$z$ galaxies have ionization
parameters significantly higher than local galaxies.
The SDSS galaxies typically show \qion\ $\sim 2\times 10^7$\,\cms,
while the high-$z$ LBGs exhibit an ionization parameter higher than 
the SDSS galaxies by a factor of $\sim 4$
(Table \ref{tbl:Z_qion}, Figure \ref{fig:Z_qion}).
This evolution of ionization parameter is consistent with the trend 
that high-$z$ galaxies depart from the local star-forming sequence
in the BPT diagram towards higher \OIII$/$\Hb\ ratio 
(e.g., \citealt{brinchmann2008,kewley2013a,kewley2013b}), 
and with the recent spectroscopic surveys of high-$z$ galaxies 
\citep{cullen2014,holden2014}.
The evolution of ionization parameter can be closely related to the 
evolution of SFR and $M_{\star}$. 
As shown in Figure \ref{fig:o3o2r23_all}, a high \OIII$/$\OII\ ratio,
i.e., a high ionization parameter, is found in low $M_{\star}$, low
$\mu_{0.32}$, and high sSFR galaxies.
Since high-$z$ galaxies tend to have a higher sSFR (or smaller $\mu_{0.32}$)
than local galaxies (Figure \ref{fig:o3o2_properties}; 
cf. the evolution of the star-formation main sequence; \citealt{daddi2007}), 
they are supposed to show an ionization parameter 
typically higher than local galaxies.
Note that ionization parameter is, by definition, the ratio of 
ionizing photon and hydrogen atom densities. Since the amount of
ionizing photons and hydrogen atoms would be positively correlated 
with SFR and stellar mass, respectively, the evolution of ionization 
parameter would be explained by the increase of sSFR, i.e., the high 
SFR and low mass for high-$z$ galaxies.

\begin{figure}
  \centerline{
    \includegraphics[width=0.75\columnwidth]{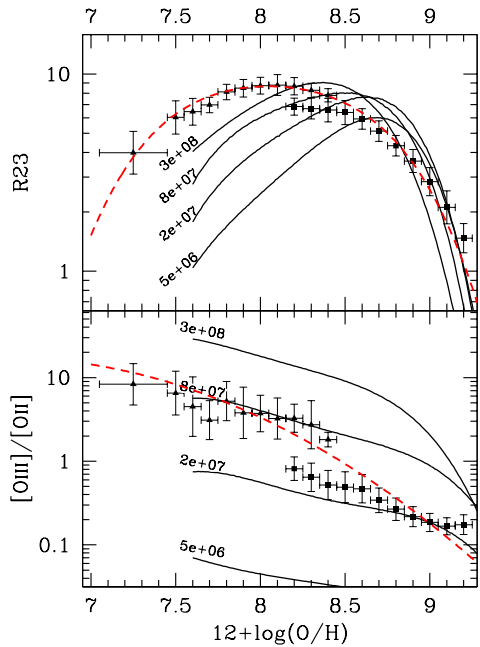}
  }
  \caption{
    $R23$-index (top) and \OIII$/$\OII\ ratio (bottom) as a function
    of metallicity.
    The triangles and squares show local galaxies in the samples (i) and
    (ii), respectively (see text), and the red dashed curve represents
    the local empirical relations \citep{maiolino2008}.
    The black solid curves denote photoionization models
    \citep{KD2002}. The labels on the curves present values of 
    $\log($\qion$)$.
  }
  \label{fig:nagao06_lines_Z}
\end{figure}

The local extreme populations of GPs, LyC leakers, and LBAs have an 
ionization parameter as high as $z\sim 2-3$ LBGs and LAEs. Moreover, 
the ionization parameter of GPs is comparable to that of LAEs. 
GP's low metallicities (see also \citealt{amorin2010}) and SFR 
\citep{izotov2011} are also analogous to those observed in LAEs.
In these senses, GPs could be local counterparts of $z\sim 2-3$ LAEs. 
However, GPs are quite rare objects in the local universe, occupying 
only $\sim 0.06$\,\% of the SDSS sample 
\citep{nakajima2013,cardamone2009}. The number density of GPs is
approximately two orders of magnitude smaller than that of $z\sim 2$ 
LAEs. Interestingly, \citet{heckman2005} report a similar difference 
of abundances between LBAs and $z\sim 3$ LBGs.
These results indicate that galaxies with a high ionization parameter
would emerge in a high redshift universe, and that these galaxies are 
more dominant in high-$z$ than the local universe.

\begin{figure*}
  \centerline{
    \includegraphics[width=0.95\textwidth]{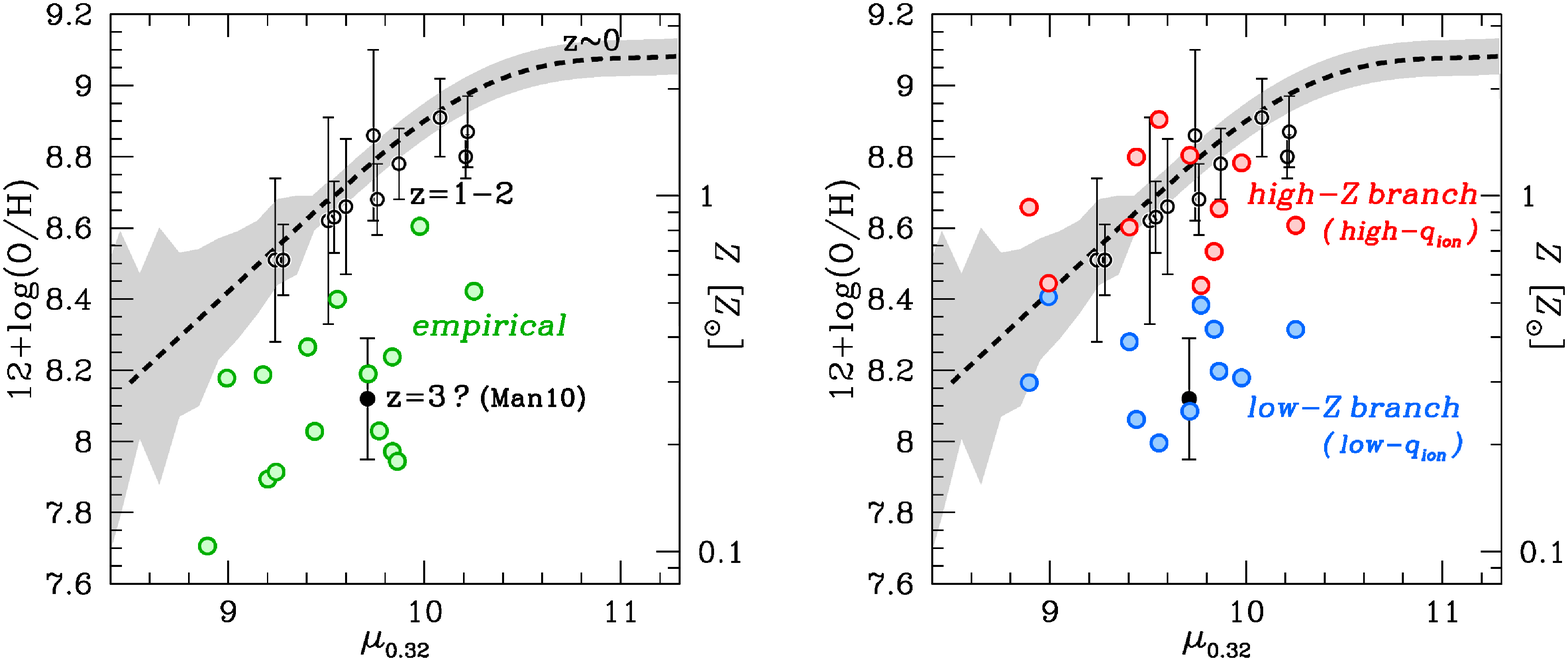}
  }
  \caption{
    $z\sim 3$ galaxies in the FMR plot.
    {\em Left:} The black dashed curve and the gray shaded area
    indicate the FMR and its typical dispersion, respectively
    \citep{mannucci2010,mannucci2011}.
    The black open and filled circles represent galaxies at $z=1$--$2$ 
    and $z\sim 3$, respectively, that are compiled by 
    \citet{mannucci2010}.
    The green circles denote $z\sim 3$ galaxies
    \citep{maiolino2008,mannucci2009} whose metallicities are estimated
    with the local empirical relations in the same manner as
    \citet{mannucci2010}. These green circles are consistent with the 
    average value shown in \citet{mannucci2010}.
    {\em Right:} Same as the left panel, but for metallicity estimates 
    using the \citet{KK2004} method that is free from the bias of 
    ionization parameters.
    The red and blue circles denote metallicities of the $z\sim 3$
    galaxies in the cases of high- and low-$Z$ branches, respectively.
    Four $z\sim 3$ galaxies are omitted from the plot, since their
    $R23$-index values are out of the range where the \citet{KK2004}
    method is available.
    The blue circles are consistent with the \citet{mannucci2010}'s 
    average value, while the red circles appear to follow the FMR.
  }
  \label{fig:fmr_z3}
\end{figure*}

Recently, some studies have found galaxies with very strong nebular 
emission lines that are called extreme emission line galaxies
(EELGs; e.g., \citealt{vanderwel2011,atek2011}).
\citet{vanderwel2011} investigate EELGs at $z\sim 1.7$ in the CANDELS 
deep imaging data of HST/WFC3. These EELGs present a very strong \OIII\ 
of EW $\ga 1000$\,\AA\ that significantly boosts flux measurements in 
a NIR broadband, which are similar to the bright $r$-band fluxes of 
GPs. Although no estimates of ionization parameter are given for EELGs
due to lacks of \OII\ data, they could have a very high ionization
parameter suggested from their very large EW(\OIII), low $M_{\star}$
($\sim 10^8\,M_{\odot}$), and high sSFR ($\sim 50$\,Gyr$^{-1}$).
\citet{kakazu2007} study ultra-strong line galaxies (USELs) at 
$z\sim 1$ identified by narrowband imaging, and find that USELs have 
extremely low metallicities (\Oabundance\ $\la 8$) and very high 
\OIII$/$\OII\ ratios ($\ga 3-100$). They argue that USELs typically 
have ionization parameters as high as \qion $\ga 10^8$\,\cms.
Obviously, ionization parameter is a key quantity to characterize
these extreme populations of galaxies.

\subsection{Revisiting the Issue of FMR Evolution from $z\sim 2$ to $3$}
\label{ssec:fmr_z3}

\citet{mannucci2010} have suggested that the FMR can describe properties
of galaxies with no redshift evolution up to $z\sim 2.5$, and claimed 
that galaxies at $z\ga 3$ fall significantly below the FMR by 
$\sim 0.6$\,dex. The analysis of \citet{mannucci2010} indicates the 
evolution of FMR from $z\sim 2$ to $3$, but it is not clear why 
the FMR evolves only between $z\sim 2$ and $z\sim 3$.
In contrast to the FMR evolution, we have argued in Section
\ref{ssec:o3o2r23_FIR} that no significant differences are found 
between galaxies at $z\sim 2$ and $3$ in the FIR
(Figure \ref{fig:o3o2r23_mu_f_solo}).
We address the question why the FMR shows the evolution but no 
evolution is found in the FIR.

\citet{mannucci2010} use metallicities of $z\ga 3$ galaxies given by 
\citet{maiolino2008} and \citet{mannucci2009} that are estimated with 
the local empirical relations of \citet{maiolino2008}.
The local empirical relations of \citet{maiolino2008} are obtained 
from two samples,
(i) $259$ local galaxies with \Oabundance\ $<8.4$ whose metallicities
are estimated by the electron temperature $T_{e}$ method, and 
(ii) $22,482$ star-forming SDSS galaxies with \Oabundance\ $>8.2$
given with the photoionization model
(\citealt{KD2002}; see also \citealt{nagao2006}).
Figure \ref{fig:nagao06_lines_Z} presents $R23$-index and \OIII$/$\OII\ 
ratio of the two samples. The red dashed curves denote the local 
empirical relations determined with these two samples 
\citep{maiolino2008}.
Figure \ref{fig:nagao06_lines_Z} compares the photoionization models 
(black solid curves) with the local empirical relations, and indicates 
that the local empirical relations implicitly assume an ionization 
parameter. Given the fact that high-$z$ galaxies have an ionization 
parameter higher than local galaxies 
(Section \ref{ssec:qion_evolution}), the local empirical relations 
assuming a relatively low ionization parameter would provide biased 
metallicity estimates for high-$z$ galaxies.
At $z\la 2.5$, the \NII$\lambda 6584$$/$\Ha\ ratio can be used to
estimate metallicities and/or to discriminate one from two metallicity 
solutions given by $R23$-index. However, at $z\ga 2.5$ 
\NII$\lambda 6584$ and \Ha\ lines cannot be detected from the ground. 
The \OIII$/$\OII\ ratio is needed to determine one from two metallicity 
solutions of $R23$-index.
The bottom panel of Figure \ref{fig:nagao06_lines_Z} shows that the 
photoionization model predicts a lower metallicity if a lower 
ionization parameter is assumed. Thus, one may underestimate 
metallicities of $z\ga 3$ galaxies with the local empirical relations, 
since the ionization parameter of $z\ga 3$ galaxies is significantly 
higher than those of local galaxies 
(Section \ref{ssec:qion_evolution}).

We evaluate the bias of $z\sim 3$ galaxy metallicity estimates,
which is originated from the local empirical relation. We estimate 
metallicities of $z\sim 3$ galaxies with the \citet{KK2004} method in 
the same manner as Section \ref{ssec:interpretation_o3o2r23}, 
allowing the ionization parameter evolution.
We compile a total of $15$ $z\sim 3$ galaxies from AMAZE 
\citep{maiolino2008} and LSD \citep{mannucci2009} projects.
First, for a consistency check, we estimate metallicities of our $15$ 
$z\sim 3$ galaxies with the local empirical relation used by
\citet{mannucci2010}. We find that the average metallicity of the 
$z\sim 3$ galaxies is \Oabundance\ $\sim 8.1$ (left panel of Figure
\ref{fig:fmr_z3}), which is consistent with that of 
\citet{mannucci2010} (\Oabundance\ $\sim 8.0-8.2$). We confirm that 
our $15$ galaxies and the local empirical relation can reproduce the 
results of \citet{mannucci2010}. Then, using the \citet{KK2004} method,
we obtain average metallicities of our $11$ $z\sim 3$ galaxies for
the low-$Z$ and high-$Z$ branches that are \Oabundance\ $=8.20$ and
$8.65$, with ionization parameters of \qion\ $\sim 6.6\times 10^7$
and $1.1\times 10^8$\,\cms, respectively. Typically $\sim 0.2-0.3$\,dex 
errors are associated with the estimates of metallicity and ionization 
parameter for each of the galaxies.
Note that four out of fifteen galaxies are omitted from these 
metallicity estimates, since their $R23$-index values fall out of the 
range covered by Equations (\ref{eq:Z_l_KK04}) and (\ref{eq:Z_h_KK04}).
The right panel of Figure \ref{fig:fmr_z3} compares the metallicity 
estimate of \citet{mannucci2010} with ours allowing the ionization 
parameter evolution. The high-$Z$ branch metallicities appear to follow 
the FMR (black dashed curve; \citealt{mannucci2010,mannucci2011}).
In contrast, the low-$Z$ branch metallicities fall below the FMR,
which is consistent with the result of \citet{mannucci2010}.

\begin{figure}
  \centerline{
    \includegraphics[width=0.95\columnwidth]{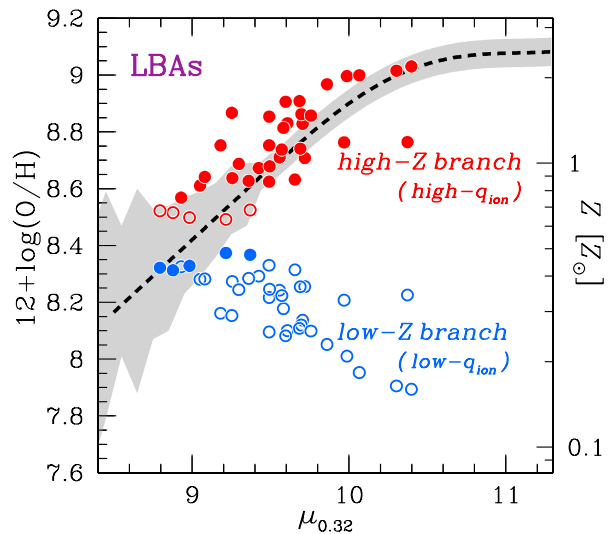}
  }
  \caption{
    LBAs in the FMR plot.
    Metallicities are estimated by the \citet{KK2004} method. 
    The red and blue circles denote metallicities 
    in the cases of high- and low-$Z$ branches, respectively.
    For each LBA, the two metallicity solutions are distinguished 
    by the other metallicity indicators (see text).
    The filled circles present metallicities that agree 
    better with the other metallicity estimates. 
    The open circles show less likely solutions.
    LBAs on average have metallicities of high-$Z$ branch
    and follow the FMR.
  }
  \label{fig:fmr_LBAs}
\end{figure}

Because there is no useful line ratio, such as \NII$\lambda 6584$$/$\Ha,
that determines one from two metallicity solutions of low and high-$Z$ 
branches, we cannot conclude which metallicities are correct and 
whether $z\sim 3$ galaxies follow the FMR.
However, we find that the high ionization parameter of 
$\sim 1.1\times 10^8$\,\cms\ for the high-$Z$ branch solution agrees 
with the our findings of ionization parameter evolution from $z=0$ to 
$3$ (Section \ref{ssec:qion_evolution}). Moreover, there are no physical 
reasons for the sudden departure from the FMR from $z=2$ to $3$, e.g., 
the smooth evolution of cosmic SFR/stellar mass densities at $z\ga 2$ 
\citep{bouwens2011,stark2013}. 
Theoretical studies do not find the departure from the FMR at $z\sim 3$.
Our metallicity estimates of high-$Z$ branch of $z\sim 3$ galaxies are 
comparable with those predicted by \citet{dave2011} who perform 
cosmological hydrodynamical simulations. \citet{dayal2013} have 
explained the trend of the FMR with the redshift-independent simple
analytic model.
These pieces of evidence imply that the $z\sim 3$ galaxy metallicities 
of high-$Z$ branch with the high-ionization parameter are correct, and 
that $z\sim 3$ galaxies follow the FMR.

As local counterparts of high-$z$ galaxies, we also examine 
LBAs in the FMR plot 
in Figure \ref{fig:fmr_LBAs}.
Their metallicities are estimated in the same manner as those
for $z\sim 3$ galaxies. However, their two metallicity solutions 
can be distinguished, since other metallicity indicators of
gas temperature and/or \NII$\lambda 6584$$/$\Ha\ ratio are 
available for the local galaxies. 
The filled circles display metallicities that agree better with 
those estimated by the other metallicity indicators.
Figure \ref{fig:fmr_LBAs} suggests that LBAs follow the FMR if 
their metallicities are estimated with ionization parameter. 
We have also checked that the FMR is valid for GPs, whose 
metallicities are estimated in the same manner as those for LBAs.
These findings confirm the validity of the FMR over the wide ranges 
of stellar mass and SFR in the local universe.
Another notable trend in Figure \ref{fig:fmr_LBAs} is that 
the LBAs metallicities of high-$Z$ branch with high-ionization 
parameter are generally correct. This is what we argue for 
the $z\sim 3$ galaxies. 
Since LBAs are local UV-selected galaxies whose characteristics 
are similar to those for LBGs (e.g., \citealt{overzier2009}), 
their trends in the FMR plot support our argument that $z\sim 3$ 
LBGs have metallicities of high-$Z$ branch with high-ionization 
parameter and follow the FMR.

For more conclusive discussion of $z\sim 3$ galaxies, 
one needs deeper NIR spectroscopy 
to provide an ionization parameter indicator of 
\NeIII$\lambda 3869$$/$\OII$\lambda 3727$ 
(e.g., \citealt{nagao2006,LR2013,richardson2013})
or IR ($\lambda>3\mu$m) spectroscopy from the space to obtain 
\NII$\lambda 6584$$/$\Ha\ that determines one from two metallicity 
solutions.
In particular, the \NeIII$/$\OII\ ratio is observable from
the ground up to $z\sim 5$. 
We have confirmed the utility of the \NeIII$/$\OII\ ratio for 
LAEs, whose ionization parameters are very high
(\citealt{fosbury2003,christensen2012b}).

\subsection{Fundamental Ionization Relation}
\label{ssec:discussion_FIR}

As demonstrated in Section \ref{ssec:fmr_z3}, ionization parameter is
important for metallicity estimates. Although the FMR reproduces 
properties of galaxies with no significant evolution, input metallicities 
to the FMR would be sometimes biased due to the evolution of ionization 
parameters.
In this sense, the FIR is advantageous because the FIR allows the 
difference of ionization parameter.
The FIR consisting of the fiducial $\mu_{0.32}$ parameter requires 
non-zero $\beta$, where $\beta = 0$ is ruled out at the $>95$\,\% 
confidence level (Section \ref{ssec:o3o2r23_FIR}). This indicates that 
the four-dimensional relation of FIR between ionization parameter,
metallicity, SFR, and $M_{\star}$ is useful to characterize gas 
properties of galaxies.

\subsection{Relationship between Ionization State and Ionizing Photon Escape}
\label{ssec:qion_fesc}

We have found in Figure \ref{fig:o3o2r23_solo} that the $z\sim 0$ LyC
leakers show \OIII$/$\OII\ ratios significantly higher than typical
low-$z$ galaxies, which suggest that these LyC leakers have a high
ionization parameter. The LyC leakers' \OIII$/$\OII\ ratios are 
comparable to those of GPs, LBAs, and $z=2-3$ galaxies.
These results imply that galaxies with a high \OIII$/$\OII\ ratio,
GPs, LBAs, and $z=2-3$ galaxies, may have a high \fesc.

\begin{figure}
  \centerline{
    \includegraphics[width=0.95\columnwidth]{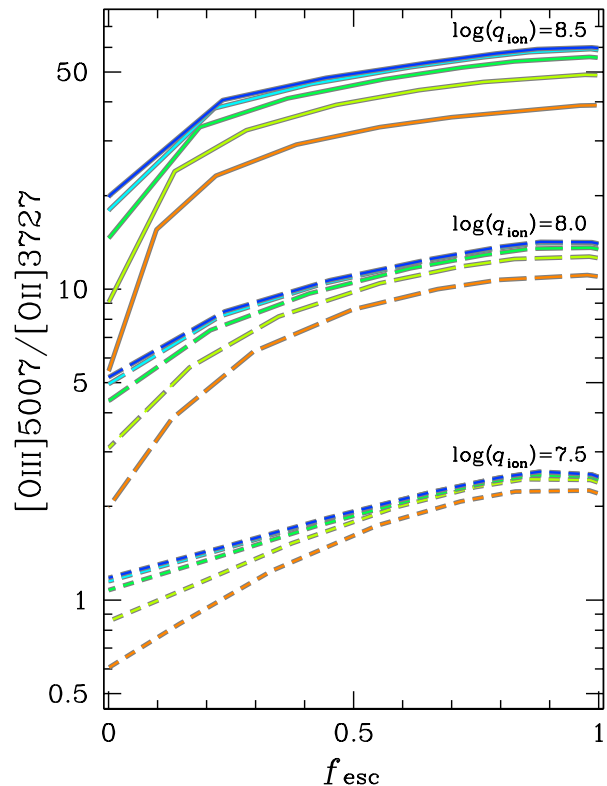}
  }
  \caption{
    \OIII$/$\OII\ vs. \fesc\ predicted by our \textsc{cloudy} models
    with $\log($\qion$)=8.5$ (solid lines), $8.0$ (long-dashed lines), 
    and $7.5$ (dashed lines). The blue, cyan, green, light green, and 
    orange lines denote \OIII$/$\OII-\fesc\ relations for a metallicity of 
    $Z=0.05$, $0.1$, $0.2$, $0.5$, and $1.0\,Z_{\odot}$, respectively.
  }
  \label{fig:o3o2_fesc}
\end{figure}

To examine the dependence of \OIII$/$\OII\ ratio on \fesc, we perform
photoionization model calculations with \textsc{cloudy}
(version 13.02; \citealt{ferland1998,ferland2013}). 
We assume a constant-density homogeneous inter-stellar gas cloud 
with a spherically closed geometry. 
A volume filling factor of the ionized gas is assumed to be unity for 
simplification.
We include dust physics and the depletion factors of the various elements 
from the gaseous phase in the same manner as the analysis of 
\citet{dopita2006b}.
In non-solar metallicities, we assume that both the dust model and the
depletion factors are unchanged, but the dust abundance is assumed to
scale linearly with the gas metallicity, as adopted in 
\citet{nagao2011}. All elements except for nitrogen, carbon, and helium 
are taken to be primary nucleosynthesis elements. For nitrogen, we use 
a form given by \citet{lopez-sanchez2012} to take account of its 
secondary nucleosynthesis component in a high metallicity range. For 
carbon and helium, we use forms in \citet{dopita2006b}.
Photoionization models are calculated with the parameters of
$Z=(0.05,\, 0.1,\, 0.2,\, 0.5,\, 1.0)\,Z_{\odot}$,
$\log($\qion$/$\cms$) =(7.5,\, 8.0,\, 8.5)$, and
$n_{\rm H}=10^2$\,cm$^{-3}$
\footnote{
We have checked that changing the density to $n_{\rm H}=10$\,cm$^{-3}$ 
and $10^{3}$\,cm$^{-3}$ does not alter our results. 
Since both \OIII\ and \OII\ are collisionally 
excited lines, their intensities are similarly enhanced 
if the density is smaller than their critical densities of 
$7\times 10^5$\,cm$^{-3}$ for \OIII\ and 
$(3$--$16)\times 10^3$\,cm$^{-3}$ for \OII\ \citep{osterbrock1989}.
\fesc\ is mainly governed by \NHI.
}.
The incident ionizing radiation fields are generated by 
\textsc{Starburst99} \citep{leitherer1999}. We use instantaneous 
burst models at zero age with a \citet{kroupa2001} IMF and lower 
and upper mass limits of $0.1$ and $120\,M_{\odot}$, respectively. 
Stellar metallicities are matched to the gas-phase ones.
Under these assumptions and conditions, we evaluate \fesc\ with 
the number of ionizing photons escaping from a thin cloud with a 
low column density. We adopt a neutral hydrogen column density, 
\NHI, as the stopping criterion of our calculations that ranges 
from \NHI\ $=10^{15}$ to $10^{20}$\,cm$^{-2}$.
The escape fraction is defined as a ratio of transmitted ionizing 
photons ($\lambda<912$\,\AA) to input ionizing photons.

\begin{figure}
  \centerline{
    \includegraphics[width=0.95\columnwidth]{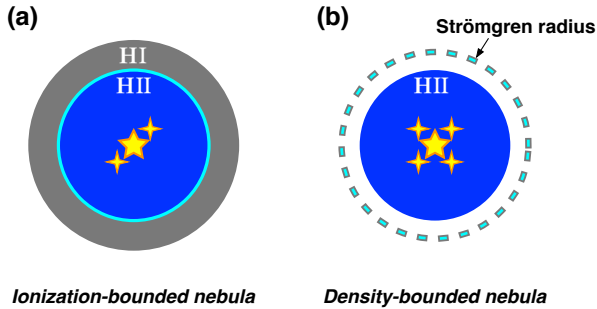}
  }
  \caption[]
          {
            Schematic illustrations of two \HII\ regions.
            A photoionized \HII\ region is presented with blue, and 
            the outer \HI\ region is shown with gray. Yellow stars are 
            central ionizing sources.
            (a) An ionization-bounded nebula whose radius is determined
            by the ionization equilibrium. The cyan shows the 
            \Stromgren\ radius.
            (b) A density-bounded nebula whose radius is determined 
            by the distribution of gas cloud. The surrounding \HI\ cloud
            is small enough that the central sources can ionized it 
            completely. The nebula thus cannot form a complete 
            \Stromgren\ sphere.  
          }
          \label{fig:HII_regions}
\end{figure}

Figure \ref{fig:o3o2_fesc} shows the \textsc{cloudy} calculation results
on the \OIII$/$\OII\ vs. \fesc\ plot.
Figure \ref{fig:o3o2_fesc} indicates that the \OIII$/$\OII\ ratio 
increases with \fesc. This trend qualitatively explains the large 
\OIII$/$\OII\ ratio of the LyC leakers in 
Figure \ref{fig:o3o2r23_solo}.
This relationship between \OIII$/$\OII\ ratio and \fesc\ is understood 
by one of the two classifications of star-forming nebulae;
\begin{enumerate}
  \renewcommand{\theenumi}{(\arabic{enumi})}
  \item Ionization-bounded nebula whose radius is determined by the 
    ionization equilibrium between the ionizing photon production rate and
    the recombination rate 
    (Equation \ref{eq:equilibrium}; Figure \ref{fig:HII_regions}a). 
    The radius of 
    the ionized region is called the \Stromgren\ radius.
  \item Density-bounded nebula whose radius is determined by the 
    distribution of gas cloud (Figure \ref{fig:HII_regions}b). 
    The density-bounded nebula does not expend
    all of ionizing photons, but leak ionizing photons that are not used 
    for the ionization of the nebula.
\end{enumerate}
The ionizing photon escape takes place in the density-bounded nebula 
under the assumption of homogeneous gas cloud.
A density-bounded nebula with a less \NHI\ emits more 
ionizing photons, and has a higher \fesc\ \citep{giammanco2005}.
On the other hand, in the ionization-bounded nebula, lower ionization 
species of O$^{+}$ (producing \OII) dominate in the outer region of 
the nebula, while an O$^{2+}$ zone (producing \OIII) exists near the 
ionizing source (e.g., \citealt{shields1990,OK1997,pellegrini2012}). 
Thus, 
the O$^{+}$ zone of density-bounded nebula is smaller than that of 
an ionization-bounded nebula, but the size of O$^{2+}$ zone is similar
in the density-bounded and ionization-bounded nebulae. As a result, 
the \OIII$/$\OII\ ratio is large in density-bounded nebulae, and 
nebulae with a less \NHI\ have a higher \OIII$/$\OII\ 
ratio.
From the combination of \NHI\ vs. \fesc\ and 
\NHI\ vs. \OIII$/$\OII\ relations, we understand the 
positive correlation between \OIII$/$\OII\ and \fesc.
We note that we consider a very simple case, with the 
assumption that \HII-regions are homogeneous density-bounded 
nebulae. In reality, conditions of \HII-regions with ionizing
photon escape could be more complex, e.g., ionization-bounded
nebulae with highly ionized low density holes 
(e.g., \citealt{zackrisson2013}; see also the end of this 
section). However, we present in this paper the simple case
to investigate the possibility that the \OIII$/$\OII\ 
ratio could be affected by the condition of \HII-regions 
where ionizing photon escapes take place.
It is beyond the scope of this paper whether or not the 
assumption of the density-bounded nebulae is reasonable.
The assumption remains to be tested by future observations.

In Figure \ref{fig:o3o2_fesc}, \OIII$/$\OII\ ratios are determined 
not only by \fesc, but also by ionization parameter and metallicity. 
For example, the \OIII$/$\OII\ ratio increases by an order of magnitude 
from $\log($\qion$)=7.5$ to $8.5$. Similarly, the \OIII$/$\OII\ ratio 
increases by a factor of $\sim 2-4$ from $Z=1.0$ and $0.1\,Z_{\odot}$, 
although this dependence is small (Section \ref{ssec:origin_o3o2}).
On the other hand, the \OIII$/$\OII\ ratio increases by a factor of 
$\sim 2-3$ from \fesc\ $=0$ to $0.5$.
The \OIII$/$\OII\ ratio is thus sensitive to \fesc\ as well as 
ionization parameter and metallicity.
These results would suggest that a galaxy with a high \OIII$/$\OII\ 
ratio can be a candidate of high \fesc\ object, but that not all of 
high \OIII$/$\OII\ objects are LyC leakers.

In Figure \ref{fig:o3o2r23_mu_f_solo}, we have found that most 
of galaxies at $z=0-3$ follow the FIR within the uncertainties.
However, some galaxies clearly depart from the FIR at a few sigma
level. The two $z\sim 0$ LyC leakers have an \OIII$/$\OII\ ratio 
higher than the FIR by a factor of $\sim 2-3$ at high 
significance. 
The enhancement of \OIII$/$\OII\ ratio may be caused by an 
ionizing photon escape. If it is true, some GPs and LAEs lying 
above the FIR would be good candidates of ionizing photon 
emitting objects.
In Figure \ref{fig:o3o2r23_mu_f_solo}, LAEs particularly have a 
very high \OIII$/$\OII\ ratio, some of which are departing from 
the FIR more significantly than LBGs over the sizes of error bars.
This physical characteristics would be originated from a very 
large \fesc\ given by a low \NHI\ of LAE. 
The low \NHI\ of LAE is also supported by the 
gas-dynamics study of \citet{hashimoto2013} who claim that LAEs 
typically have a \NHI\ lower than LBGs from the analysis 
of velocity offsets between a \Lya\ line and low-ionization 
interstellar absorption lines with respect to the systemic 
velocity.
This tendency is confirmed with a larger number of LAEs 
by \citet{shibuya2014a}.
In this companion paper, \citet{shibuya2014b} examine LAE structures,
suggesting that LAEs with a large \Lya\ EW tend to have a
small ellipticity. This is consistent with the theoretical results
that \Lya\ photons can more easily escape from face-on disks 
having a small ellipticity, due to a low \NHI\ 
(e.g., \citealt{verhamme2012,yajima2012,ZW2013}).
Moreover, recent deep narrowband imaging surveys have indicated 
diffuse \Lya\ emitting halos around high-$z$ star-forming galaxies 
that are more prominent for LBGs (e.g., \citealt{steidel2011}) 
than LAEs \citep{feldmeier2013}.
Since most of the observed \Lya\ emission in the diffuse halos 
is probably originated from the galaxy \HII-regions and 
scattered to the line of sight by extended \HI\ gas in the 
galaxy's circum-galactic medium (CGM), LAEs could exhibit \Lya\ 
halos weaker than LBGs due to their smaller amounts of \HI\ gas
in the CGM. 
This scenario is consistent with our argument of low \NHI\ of LAE. 
Interestingly, that interpretation also consistently explains 
the results of direct LyC detections of $z\sim 3$ galaxies
that reveal a high \fesc\ for LAEs ($\sim 0.1-0.3$) than
LBGs ($\sim 0.05$) \citep{iwata2009,nestor2011,nestor2013}.
Thus, an origin of the high \fesc\ in LAEs would be their 
optically thin \HI\ gas surrounding star-forming regions.
A similar suggestion of the high \fesc\ is inferred for GPs
\citep{JO2013}.
The local LyC leakers show an \fesc\ comparably high as 
LBGs ($\sim 0.02-0.1$; \citealt{bergvall2006,leitet2013}),
and thus have an \OIII$/$\OII\ ratio as high as LBGs.

LAEs with a high \fesc\ would play a key role in supplying 
ionizing photons for cosmic reionization in the early universe.
Previous $z\sim 6-7$ galaxy surveys have suggested that the
universe could not be totally ionized by galaxies alone at
$z\sim 6-7$ if one assumes similar properties of galaxies at
$z\sim 3$ that include \fesc\ $\sim 0.05$
(e.g., \citealt{ouchi2009,robertson2010}).
Because the fraction of LAEs in star-forming galaxies increases 
with redshift (at least up to $z\sim 6$; 
e.g., \citealt{ouchi2008,vanzella2009,stark2010,stark2011,hayes2011}), 
the average \fesc\ for $z\sim 6$ galaxies would be higher 
than that for $z\sim 3$ galaxies. The higher \fesc\ for $z\sim 6$ 
galaxies may resolve the problem of ionizing photon 
deficit in cosmic reionization.
Note that the number of observed LAEs suddenly drops from 
$z\sim 6$ to $7$.
This could be due to the \Lya\ damping absorption of partly 
neutral IGM 
\citep{vanzella2011,pentericci2011,ono2012,schenker2012}
or optically thick systems in the intervening IGM 
\citep{BH2013}.
Alternatively, the decreasing fraction of LAEs could be
explained by the intrinsic decrease of galaxies with 
strong \Lya\ emission at $z>6$ \citep{dijkstra2014}.
If we assume the former scenarios, galaxies with an 
intrinsically bright \Lya\ line would not decrease from 
$z\sim 6$ to $7$.
However, this is still an open question.
A similar implication is provided by \citet{schenker2013} on the
basis of a small velocity offset of \Lya\ line with respect to 
the systemic velocity for $z=3-4$ galaxies.
We also note that this picture would be consistent with the 
observational fact that gas fraction of star-forming galaxies 
increases with redshift, or more properly, with sSFR 
(e.g., \citealt{tacconi2010,tacconi2013,santini2014}).
If the molecular gas exists in dense cores of clumps embedded in 
density-bounded nebulae ionized by young stars, the galaxy could 
have a high gas fraction and high \fesc. 
A dense molecular gas clump could reside inside the \HII-region,
since the dense gas could be self-shielded from ionizing radiation.
A good example is found in the Orion nebula 
(see, e.g., \citealt{GS1989} for a review).
Such clumpy geometry of dense molecular gas could explain the 
high gas fraction, active star-formation, and high \fesc\ in 
high-$z$ LAEs.
\\

\begin{figure}
  \centerline{
    \includegraphics[width=0.95\columnwidth]{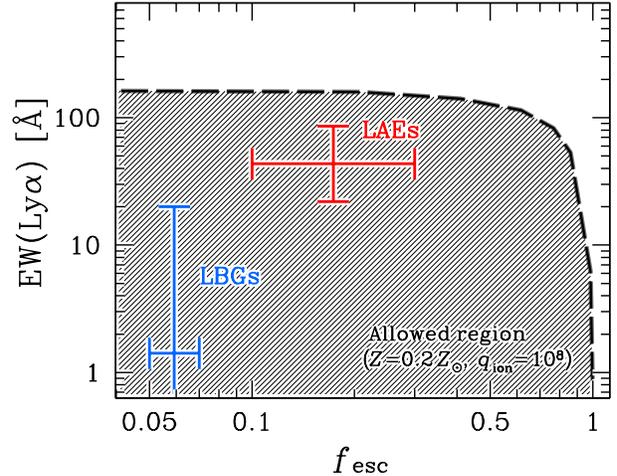}
  }
  \caption{
    EW(\Lya) vs. \fesc\ plot. 
    The dashed curve is our \textsc{cloudy} model
    for $Z=0.2\,Z_{\odot}$ and $\log($\qion$)=8.0$ that
    represents upper limits of EW(\Lya) values (see text).
    Observational results by \citet{nestor2013}
    for $z\sim 3$ LAEs and LBGs are presented with red and blue colors, 
    respectively.
  }
  \label{fig:ewLya_fesc_nestor13}
\end{figure}

In the discussion of high \fesc\ for LAEs, there is one caveat that 
\Lya\ emission would become weaker in a high \fesc, because less 
ionizing photons are spent for making ionized gas that emit \Lya.
Therefore, LAEs may not be a population of high \fesc.
The dashed curve in Figure \ref{fig:ewLya_fesc_nestor13} shows 
the relation between EW(\Lya) and 
\fesc\ predicted by our \textsc{cloudy} models.
Here we assume \qion\ $=10^{8}$\,\cms\ and $Z=0.2\,Z_{\odot}$, typical 
values observed in GPs \citep{amorin2010} and $z\sim 2$ LAEs 
(\citealt{nakajima2013}).
The dashed curve presents that the EW(\Lya) decreases
sharply if its \fesc\ are over $\simeq 0.8$.
However, in \fesc\ $\simeq 0.0-0.8$, EW(\Lya) remains unchanged 
within a factor of $\la 2$.
We have checked that the overall trends are the same, adopting other 
values of \qion\ and $Z$. For example, EW(\Lya) increases (decreases)
by a factor of $\sim 2-3$ for $Z=0.05\,Z_{\odot}$ ($1\,Z_{\odot}$).
LAEs are usually defined as galaxies with EW(\Lya) $>20$\,\AA.
If an escape fraction is very high, \fesc\ $\ga 0.9$, a galaxy cannot 
be observed as an LAE. However, LAEs, 
or galaxies with EW(\Lya) $>20$\,\AA, can exist in the range of 
\fesc\ $\la 0.8$.
Since a very high \fesc\ of $\ga 0.8$ is not required for cosmic 
reionization but only \fesc\ $\simeq 0.2$ \citep{robertson2013},
the weakening of Ly$\alpha$ in density-bounded nebulae does not 
rule out the possibility that major ionizing sources are LAEs.
Figure \ref{fig:ewLya_fesc_nestor13} also compares our model and the 
recent observational results for $z\sim 3$ LAEs and LBGs whose
\fesc\ are measured from LyC fluxes \citep{nestor2013}%
\footnote{
We refer to \citet{nestor2011} and \citet{shapley2003} for the 
ranges of EW(\Lya) of LAEs and LBGs, respectively. 
}.
EW(\Lya) values of our model are regarded as upper limits, 
because EW(\Lya) decreases by dust extinction and hydrogen 
scattering. The observational results fall below our model, 
and are consistent with our model. The observations indicate
a positive correlation between \fesc\ and EW(\Lya) in 
\fesc\ $\la 0.5$, while our model shows a nearly constant 
EW(\Lya) in this regime. This difference would infer that 
LAEs have a less dust extinction, a more homogeneous geometry
for isotropic \Lya\ scattering, a less metallicity, and 
more ionizing photons than LBGs.

We note that our calculations assume a density-bounded nebula with a
spherically closed geometry, and predict dependencies of 
\NHI\ on the emergent spectra. A volume filling factor of 
the ionized gas is assumed to be unity. 
Thus, we have presented this specific case of EW(\Lya) as a function 
of \fesc. However, the EW(\Lya) and \fesc\ involve various complex 
physical effects.
EW(\Lya) depends on gas and dust geometries 
(e.g., \citealt{neufeld1991,scarlata2009}; cf. \citealt{duval2014}) 
and galactic scale outflow (e.g., \citealt{kunth1998, verhamme2006}).
\fesc\ anti-correlates with a covering fraction of neutral gas.
Recent spectroscopic studies suggest a low covering fraction of 
neutral gas in high-$z$ star-forming galaxies
(e.g., \citealt{shapley2003,steidel2010,jones2013}).
There exist the other complex physics of EW(\Lya) and \fesc.
Nevertheless, our analysis reveals the relation of EW(\Lya) and 
\fesc\ for the case that \HII-regions are homogeneous 
density-bounded nebulae.

\section{Summary}
\label{sec:summary}

We investigate ionization state of hot ISM in galaxies at $z=0-3$ with 
$\sim 140,000$ SDSS galaxies and $108$ intermediate and high redshift 
galaxies, using an ionization parameter and metallicity sensitive line 
ratios of \OIII$\lambda 5007$$/$\OII$\lambda 3727$ and
$($\OIII$\lambda\lambda 5007,4959$$+$\OII$\lambda 3727)$$/$\Hb\
($R23$-index).
Our main results are summarized as follows.

\begin{itemize}
\item We confirm that $z\sim 2-3$ galaxies show an \OIII$/$\OII\ ratio
  significantly higher than a typical star-forming galaxy of SDSS
  by a factor of 
  $\ga 10$ (Figure \ref{fig:o3o2r23_solo}).
  The photoionization models reveal that these high-$z$ galaxies have
  an ionization parameter of $\log($\qion$/$\cms$)\sim 7.6-9.0$, a 
  factor of $\sim 4-10$ higher than local galaxies
  (Figure \ref{fig:Z_qion}).
\item We find a high ionization parameter in local extreme populations 
  of green pea galaxies (GPs), Lyman-continuum emitting galaxies 
  (LyC leakers), and Lyman break analogs (LBAs), and the ionization
  parameters of these galaxies are comparable with the ones of $z=2-3$ 
  galaxies (Figure \ref{fig:Z_qion}).
  These local extreme populations are quite rare, and their number
  densities are nearly two orders of magnitude smaller than those of
  $z=2-3$ galaxies.
\item We identify a correlation between the \OIII$/$\OII\ ratio and
  galaxy global properties of star-formation rate (SFR), stellar mass
  ($M_{\star}$), and metallicity (Figure \ref{fig:o3o2_properties}).
  A high \OIII$/$\OII\ ratio is found in less massive,
  efficiently star-forming, and metal-poor galaxies.
\item We develop a four-dimensional relation of ionization parameter,
  SFR, $M_{\star}$, and metallicity (Figure \ref{fig:o3o2r23_all}),
  extending the fundamental metallicity relation (FMR;
  \citealt{mannucci2010,lara-lopez2010}) with ionization parameter.
  The intermediate and high-$z$ galaxies up to $z\simeq 3$ as well as 
  the local extreme populations follow the same relation defined with 
  the SDSS galaxies (Figure \ref{fig:o3o2r23_mu_f_solo}).
  The relation is thus referred to as the fundamental ionization
  relation (FIR).
\item We particularly find no significant differences between galaxies 
  at $z\sim 2$ and $3$ in the FIR (Figure \ref{fig:o3o2r23_mu_f_solo}).
  This is in contrast with the FMR whose possible evolution from
  $z\sim 2$ to $3$ is reported \citep{mannucci2010}.
  We suggest that the FMR evolution can arise, if one omits
  ionization parameter differences and adopts local empirical
  metallicity relations for high-$z$ galaxies.
  We indicate that the FMR evolution does not exist for one out of two 
  average metallicity solutions of $z\sim 3$ galaxies with a high 
  ionization parameter of $\log($\qion$/$\cms$)\ga 8$
  (right panel of Figure \ref{fig:fmr_z3}).
\item We find the $z\sim 0$ LyC leakers exhibit \OIII$/$\OII\ ratios
  significantly higher than typical low-$z$ galaxies,
  although there are only two LyC leakers.
  This trend suggests a positive correlation between \OIII$/$\OII\
  and ionizing photon escape fraction (\fesc).
  Our photoionization models of \textsc{cloudy} reproduce the trend
  (Figure \ref{fig:o3o2_fesc}).
  Since \OIII$/$\OII\ ratios for $z=2-3$ galaxies, especially
  \Lya\ emitters (LAEs), are comparable to, or higher than,
  those of the low-$z$ LyC leakers, these high-$z$ galaxies
  are candidates of high \fesc\ objects.
\item Our \textsc{cloudy} photoionization models reveal that 
  EW(\Lya) remains almost unchanged if its \fesc\ is $\simeq 0-0.8$
  (Figure \ref{fig:ewLya_fesc_nestor13}), indicating that galaxies are 
  identified as LAEs of a Ly$\alpha$ emitting population if their 
  \fesc\ values are below $\simeq 0.8$.
  Faint galaxies with intrinsically-bright Ly$\alpha$ having a high 
  \fesc\ could significantly contribute to the ionizing photon 
  production for cosmic reionization, since recent observations 
  report that a fraction of LAEs in star-forming galaxies increases 
  with redshift at the epoch with no IGM damping absorption 
  (e.g., \citealt{ouchi2008,vanzella2009,stark2010,stark2011}).
  If it is true, this would ease the tension of ionizing photon 
  deficit problem (e.g., \citealt{ouchi2009,robertson2010}).
  \\
\end{itemize}

\noindent
{\bf Acknowledgments}
We are grateful to Tohru Nagao, Lisa J. Kewley, 
Kazuhiro Shimasaku, 
Filippo Mannucci, Giovanni Cresci, Alessandro Marconi,
Eros Vanzella, Maritza Lara-Lopez, Masato Onodera, 
Jarle Brinchmann, Maryam Shirazi, Yuu Niino, 
Ken-ichi Tadaki, Sally Oey,
Kentaro Motohara, Toshikazu Shigeyama, Yuzuru Yoshii, 
Hideyuki Kobayashi, and Nobunari Kashikawa for their 
helpful comments and suggestions. 
A careful reading and valuable report by the referee is 
very much appreciated.
We thank the MPA/JHU teams which made their measured quantities on 
SDSS galaxies publicly available.
This work was supported by World Premier International Research Center
Initiative (WPI Initiative), MEXT, Japan, and KAKENHI (23244025)
Grant-in-Aid for Scientific Research (A) through Japan Society for the
Promotion of Science (JSPS). K.N. also acknowledges the JSPS Research
Fellowship for Young Scientists.




\begin{thebibliography}{}
\addcontentsline{toc}{chapter}{\bibname}
\expandafter\ifx\csname natexlab\endcsname\relax\def\natexlab#1{#1}\fi

\bibitem[\protect\citeauthoryear{Amor\'{i}n et al.}{2010}]{amorin2010} Amor\'{i}n, R.~O., P\'{e}rez-Montero, E., V\'{i}lchez, J.~M., 2010, 715, L128
\bibitem[\protect\citeauthoryear{Asplund et al.}{2009}]{asplund2009} Asplund, M., Grevesse, N., Sauval, A.~J., Scott, P., 2009, ARA\&A, 47, 481
\bibitem[\protect\citeauthoryear{Atek et al.}{2011}]{atek2011} Atek, H.,et al., 2011, \apj, 743, 121
\bibitem[\protect\citeauthoryear{Baldwin et al.}{1981}]{baldwin1981} Baldwin, J.~A., Phillips, M.~M., Terlevich, R., 1981, \pasp, 93, 5
\bibitem[\protect\citeauthoryear{Basu-Zych et al.}{2007}]{basu-zych2007} Basu-Zych, A.~R., et al., 2007, \apjs, 173, 457
\bibitem[\protect\citeauthoryear{Belli et al.}{2013}]{belli2013} Belli, S., Jones, T., Ellis, R.~S., Richard, J., 2013, \apj, 772, 141
\bibitem[\protect\citeauthoryear{Bergvall \& \"{O}stlin}{2002}]{BO2002} Bergvall, N., \"{O}stlin, G., 2002, \aap, 390, 891
\bibitem[\protect\citeauthoryear{Bergvall et al.}{2006}]{bergvall2006} Bergvall, N., Zackrisson, E., Andersson, B.~-G., Arnberg, D., Masegosa, J., \"{O}stlin, G., 2006, \aap, 448, 513
\bibitem[\protect\citeauthoryear{Bolton \& Haehnelt}{2013}]{BH2013} Bolton, J.~S., Haehnelt, M.~G., 2013, \mnras, 429, 1695
\bibitem[\protect\citeauthoryear{Bouwens et al.}{2011}]{bouwens2011} Bouwens, R.~J., et al., 2011, \apj, 737, 90
\bibitem[\protect\citeauthoryear{Brinchmann et al.}{2004}]{brinchmann2004} Brinchmann, J., Charlot, S., White, S.~D.~M., Tremonti, C., Kauffmann, G., Heckman, T., Brinkmann, J., 2004, \mnras, 351, 1151
\bibitem[\protect\citeauthoryear{Brinchmann et al.}{2008}]{brinchmann2008} Brinchmann, J., Pettini, M., Charlot, S., 2008, \mnras, 385, 769
\bibitem[\protect\citeauthoryear{Calzetti et al.}{2000}]{calzetti2000} Calzetti, D., Armus, L., Bohlin, R.~C., Kinney, A.~L., Koornneef, J., Storchi-Bergmann, T., 2000, \apj, 533, 682
\bibitem[\protect\citeauthoryear{Cardamone et al.}{2009}]{cardamone2009} Cardamone, C., et al., 2009, \mnras, 399, 1191
\bibitem[\protect\citeauthoryear{Cardelli et al.}{1989}]{cardelli1989} Cardelli, J.~A., Clayton, C., Mathis, J.~S., 1989, \apj, 345, 245
\bibitem[\protect\citeauthoryear{Chabrier}{2003}]{chabrier2003} Chabrier, G., 2003, \pasp, 115, 763
\bibitem[\protect\citeauthoryear{Charlot \& Longhetti}{2001}]{CL2001} Charlot, S., Longhetti, M., 2001, \mnras, 323, 887
\bibitem[\protect\citeauthoryear{Christensen et al.}{2012a}]{christensen2012a} Christensen, L., et al., 2012a, \mnras, 427, 1953
\bibitem[\protect\citeauthoryear{Christensen et al.}{2012b}]{christensen2012b} Christensen, L., et al., 2012b, \mnras, 427, 1973
\bibitem[\protect\citeauthoryear{Cresci et al.}{2012}]{cresci2012} Cresci, G., Mannucci, F., Sommariva, V., Maiolino, R., Marconi, A., Brusa, M., 2012, \mnras, 421, 262
\bibitem[\protect\citeauthoryear{Cullen et al.}{2014}]{cullen2014} Cullen, F., Cirasuolo, M., McLure, R.~J., Dunlop, J.~S., Bowler, R.~A.~A. 2014, \mnras, 440, 2300
\bibitem[\protect\citeauthoryear{Daddi et al.}{2007}]{daddi2007} Daddi, E., et al., 2007, \apj, 670, 156
\bibitem[\protect\citeauthoryear{Dav\'{e}}{2008}]{dave2008} Dav\'{e}, R., 2008, \mnras, 385, 147
\bibitem[\protect\citeauthoryear{Dav\'{e} et al.}{2011}]{dave2011} Dav\'{e}, R., Finlator, K., Oppenheimer, B.~D., 2011, \mnras, 416, 1354
\bibitem[\protect\citeauthoryear{Dayal et al.}{2013}]{dayal2013} Dayal, P., Ferrara, A., Dunlop, J.~S., 2013, \mnras, 430, 2891
\bibitem[\protect\citeauthoryear{Dijkstra et al.}{2014}]{dijkstra2014} Dijkstra, M., Wyithe, S., Haiman, Z., Mesinger, A., Pentericci, L., 2014, arXiv e-prints, arXiv:1401.7676
\bibitem[\protect\citeauthoryear{Dopita et al.}{2000}]{dopita2000} Dopita, M.~A., Kewley, L.~J., Heisler, C.~A., Sutherland, R.~S., 2000, \apj, 542, 224
\bibitem[\protect\citeauthoryear{Dopita et al.}{2006a}]{dopita2006a} Dopita, M.~A., et al., 2006a, \apj, 647, 244
\bibitem[\protect\citeauthoryear{Dopita et al.}{2006b}]{dopita2006b} Dopita, M.~A., et al., 2006b, \apjs, 167, 177
\bibitem[\protect\citeauthoryear{Dunkley et al.}{2009}]{dunkley2009} Dunkley, J., et al., 2009, \apjs, 180, 306
\bibitem[\protect\citeauthoryear{Duval et al.}{2014}]{duval2014} Duval, F., Schaerer, D., \"{O}stlin, G., Laursen, P., 2014, \aap, 562, 52
\bibitem[\protect\citeauthoryear{Ellison et al.}{2008}]{ellison2008} Ellison, S.~L., Patton, D.~R., Simard, L., McConnachie, A.~W., 2008, \apj, 672, L107
\bibitem[\protect\citeauthoryear{Erb et al.}{2006}]{erb2006} Erb, D.~K., Shapley, A.~E., Pettini, M., Steidel, C.~C., Reddy, N.~A., Adelberger, K.~L., 2006, \apj, 644, 813
\bibitem[\protect\citeauthoryear{Erb et al.}{2010}]{erb2010} Erb, D.~K., Pettini, M., Shapley, A.~E., Steidel, C.~C., Law, D.~R., Reddy, N.~A., 2010, \apj, 719, 1168
\bibitem[\protect\citeauthoryear{Fan et al.}{2006}]{fan2006} Fan, X., et al., 2006, \aj, 132, 117
\bibitem[\protect\citeauthoryear{Feldmeier et al.}{2013}]{feldmeier2013} Feldmeier, J.~J., et al., 2013, \apj, 776, 75
\bibitem[\protect\citeauthoryear{Ferland et al.}{1998}]{ferland1998} Ferland, G.~J., et al., 1998, \pasp, 110, 761
\bibitem[\protect\citeauthoryear{Ferland et al.}{2013}]{ferland2013} Ferland, G.~J., et al., 2013, RMxAA, 49, 137
\bibitem[\protect\citeauthoryear{Finkelstein et al.}{2009}]{finkelstein2009} Finkelstein, S.~L., et al., 2009, \apj, 700, 376
\bibitem[\protect\citeauthoryear{Fosbury et al.}{2003}]{fosbury2003} Fosbury, R.~A.~E., et al., 2003, \apj, 596, 797
\bibitem[\protect\citeauthoryear{Genzel \& Stutzki}{1989}]{GS1989} Genzel, R., Stutzki, J., 1989, ARA\&A, 27, 41
\bibitem[\protect\citeauthoryear{Giammanco et al.}{2005}]{giammanco2005} Giammanco, C., Beckman, J.~E., Cedr\'{e}s, B, 2005, \aap, 438, 599
\bibitem[\protect\citeauthoryear{Gilbank et al.}{2010}]{gilbank2010} Gilbank, D.~G., Baldry, I.~K., Balogh, M.~L., Glazebrook, K., Bower, R.~G., 2010, \mnras, 405, 2594
\bibitem[\protect\citeauthoryear{Hainline et al.}{2009}]{hainline2009} Hainline, K.~N., Shapley, A.~E., Kornei, K.~A., Pettini, M., Buckley-Geer, E., Allam, S.~S., Tucker, D.~L., 2009, \apj, 701, 52
\bibitem[\protect\citeauthoryear{Hashimoto et al.}{2013}]{hashimoto2013} Hashimoto, T., et al., 2013, \apj, 765, 70
\bibitem[\protect\citeauthoryear{Hayes et al.}{2011}]{hayes2011} Hayes, M., Schaerer, D., \"{O}stlin, G., Mas-Hesse, J.~M., Atek, H., Kunth, D., 2011, \apj, 730, 8
\bibitem[\protect\citeauthoryear{Heckman et al.}{2005}]{heckman2005} Heckman, T.~M., et al., 2005, \apjl, 619, L35
\bibitem[\protect\citeauthoryear{Heckman et al.}{2011}]{heckman2011} Heckman, T.~M., et al., 2011, \apj, 730, 5
\bibitem[\protect\citeauthoryear{Henry et al.}{2013}]{henry2013} Henry, A., Kartin, C.~L., Finlator, K., Dressler, A., 2013, \apj, 769, 148
\bibitem[\protect\citeauthoryear{Holden et al.}{2014}]{holden2014} Holden, B.~P., et al., 2014, arXiv e-prints, arXiv:1401.5490
\bibitem[\protect\citeauthoryear{Hoopes et al.}{2007}]{hoopes2007} Hoopes, C.~G., et al., 2007, \apjs, 173, 441
\bibitem[\protect\citeauthoryear{Iwata et al.}{2009}]{iwata2009} Iwata, I., et al., 2009, \apj, 692, 1287
\bibitem[\protect\citeauthoryear{Izotov et al.}{2011}]{izotov2011} Izotov, Y.~I., Guseva, N.~G., Thuan, T.~X., 2011, \apj, 728, 161
\bibitem[\protect\citeauthoryear{Jaskot \& Oey}{2013}]{JO2013} Jaskot, A.~E., Oey, M.~S., 2013, \apj, 766, 91
\bibitem[\protect\citeauthoryear{Jones et al.}{2013}]{jones2013} Jones, T.~A., Ellis, R.~S., Schenker, M.~A., Stark, D.~P., 2013, \apj, 779, 52
\bibitem[\protect\citeauthoryear{Kakazu et al.}{2007}]{kakazu2007} Kakazu, Y., Cowie, L.~L., Hu, E.~M., 2007, \apj, 668, 853
\bibitem[\protect\citeauthoryear{Kauffmann et al.}{2003a}]{kauffmann2003a} Kauffmann, et al., 2003a, \mnras, 346, 1055
\bibitem[\protect\citeauthoryear{Kauffmann et al.}{2003b}]{kauffmann2003b} Kauffmann, G., et al., 2003b, \mnras, 341, 33
\bibitem[\protect\citeauthoryear{Kennicutt}{1998}]{kennicutt1998} Kennicutt, R.~C., Jr., 1998, ARA\&A, 36, 189
\bibitem[\protect\citeauthoryear{Kewley \& Dopita}{2002}]{KD2002} Kewley, L.~J., Dopita, M.~A., 2002, \apjs, 142, 35
\bibitem[\protect\citeauthoryear{Kewley et al.}{2013a}]{kewley2013a} Kewley, L.~J., Maier, C., Yabe, K., Ohta, K., Akiyama, M., Dopita, M.~A., Yuan, T., 2013a, \apj, 774L, 10
\bibitem[\protect\citeauthoryear{Kewley et al.}{2013b}]{kewley2013b} Kewley, L.~J., et al., 2013b, \apj, 774, 100
\bibitem[\protect\citeauthoryear{Kobulnicky \& Kewley}{2004}]{KK2004} Kobulnicky, H.~A., Kewley, L.~J., 2004, \apj, 617, 240
\bibitem[\protect\citeauthoryear{Kornei et al.}{2010}]{kornei2010} Kornei, A.~K., Shapley, A.~E., Steidel, C.~C., Reddy, N.~A., Pettini, M., Bogosavljevi\'{c}, M., 2010, \apj, 711, 693
\bibitem[\protect\citeauthoryear{Kroupa}{2001}]{kroupa2001} Kroupa, P., 2001, \mnras, 322, 231
\bibitem[\protect\citeauthoryear{Kunth et al.}{1998}]{kunth1998} Kunth, D., Mas-Hesse, J.~M., Terlevich, E., Terlevich, R., Lequeux, J., Fall, S.~M., 1998, \aap, 334, 11
\bibitem[\protect\citeauthoryear{Lara-L\'{o}pez et al.}{2010}]{lara-lopez2010} Lara-L\'{o}pez, M.~A., et al., 2010, \aap, 521, L53
\bibitem[\protect\citeauthoryear{Lara-L\'{o}pez et al.}{2013}]{lara-lopez2013} Lara-L\'{o}pez, M.~A., et al., 2013, \mnras, 434, 451
\bibitem[\protect\citeauthoryear{Leitet et al.}{2011}]{leitet2011} Leitet, E., Bergvall, N., Piskunov, N. Andersson, B.~-G., 2011, \aap, 532, A107
\bibitem[\protect\citeauthoryear{Leitet et al.}{2013}]{leitet2013} Leitet, E., Bergvall, N., Hayes, M., Linn\'{e}, S., Zackrisson, E., 2013, \aap, 553, A106
\bibitem[\protect\citeauthoryear{Leitherer et al.}{1999}]{leitherer1999} Leitherer, C., et al., 1999, \apjs, 123, 3
\bibitem[\protect\citeauthoryear{Levesque \& Richardson}{2013}]{LR2013} Levesque, E.~M., Richardson, M.~L.~A., 2013, \apj, 780, 100
\bibitem[\protect\citeauthoryear{Lilly et al.}{2003}]{lilly2003} Lilly, S.~J., Carollo, C.~M., Stockton, A.~N., 2003, \apj, 597, 730
\bibitem[\protect\citeauthoryear{Liu et al.}{2008}]{liu2008} Liu, X., Shapley, A.~E., Coil, A.~L., Brinchmann, J., Ma, C.~-P., 2008, \apj, 678, 758
\bibitem[\protect\citeauthoryear{L\'{o}pez-S\'{a}nchez et al.}{2012}]{lopez-sanchez2012} L\'{o}pez-S\'{a}nchez, \'{A}.~R., Dopita, M.~A., Kewley, L.~J., Zahid, H.~J., Nicholls, D.~C., Scharw\"{a}chter, J., 2012, \mnras, 426, 2630 
\bibitem[\protect\citeauthoryear{Maier et al.}{2005}]{maier2005} Maier, C., Lilly, S.~J., Carollo, C.~M., Stockton, A., Brodwin, M., 2005, \apj, 634, 849
\bibitem[\protect\citeauthoryear{Maiolino et al.}{2008}]{maiolino2008} Maiolino, R., et al., 2008, \aap, 488, 463
\bibitem[\protect\citeauthoryear{Mannucci et al.}{2009}]{mannucci2009} Mannucci, F., et al., 2009, \mnras, 398, 1915
\bibitem[\protect\citeauthoryear{Mannucci et al.}{2010}]{mannucci2010} Mannucci, F., Cresci, G., Maiolino, R., Marconi, A., Gnerucci, A., 2010, \mnras, 408, 2115
\bibitem[\protect\citeauthoryear{Mannucci et al.}{2011}]{mannucci2011} Mannucci, F., Salvaterra, R., Campisi, M.~A., 2011, \mnras, 414, 1263
\bibitem[\protect\citeauthoryear{Nagao et al.}{2006}]{nagao2006} Nagao, T., Maiolino, R., Marconi, A., 2006, \aap, 459, 85
\bibitem[\protect\citeauthoryear{Nagao et al.}{2011}]{nagao2011} Nagao, T., Maiolino, R., Marconi, A., Matsuhara, H., 2011, \aap, 526, 149
\bibitem[\protect\citeauthoryear{Nakajima et al.}{2012}]{nakajima2012} Nakajima, et al., 2012, \apj, 745, 12
\bibitem[\protect\citeauthoryear{Nakajima et al.}{2013}]{nakajima2013} Nakajima, K., Ouchi, M., Shimasaku, K., Hashimoto, T., Ono, Y., Lee, J.~C., 2013, \apj, 769, 3
\bibitem[\protect\citeauthoryear{Narayanan \& Dav\'{e}}{2013}]{ND2013} Narayanan, D., Dav\'{e}, R., 2013, \mnras, 436, 2892
\bibitem[\protect\citeauthoryear{Nestor et al.}{2011}]{nestor2011} Nestor, D.~B., Shapley, A.~E., Steidel, C.~C., Siana, B., 2011, \apj, 736, 18
\bibitem[\protect\citeauthoryear{Nestor et al.}{2013}]{nestor2013} Nestor, D.~B., Shapley, A.~E., Kornei, K.~A., Steidel, C.~C., Siana, B., 2013, \apj, 765, 47
\bibitem[\protect\citeauthoryear{Neufeld}{1991}]{neufeld1991} Neufeld, D.~A., 1991, \apjl, 370, L85
\bibitem[\protect\citeauthoryear{Niino}{2012}]{niino2012} Niino, Y., 2012, \apj, 761, 126
\bibitem[\protect\citeauthoryear{Oey \& Kennicutt}{1997}]{OK1997} Oey, M.~S., Kennicutt, R.~C., Jr., 1997, \mnras, 291, 827
\bibitem[\protect\citeauthoryear{Ono et al.}{2012}]{ono2012} Ono Y., et al., 2012, \apj, 744, 83 
\bibitem[\protect\citeauthoryear{Osterbrock}{1989}]{osterbrock1989} Osterbrock, D.~E., 1989, Astrophysics of Gaseous Nebulae and Active Galactic Nuclei (Suasalito, CA: University Science Books)
\bibitem[\protect\citeauthoryear{Ouchi et al.}{2008}]{ouchi2008} Ouchi, M., et al., 2008, \apjs, 176, 301
\bibitem[\protect\citeauthoryear{Ouchi et al.}{2009}]{ouchi2009} Ouchi, M., et al., 2009, \apj, 706, 1136
\bibitem[\protect\citeauthoryear{Overzier et al.}{2008}]{overzier2008} Overzier, R.~A., et al., 2008, \apj, 677, 37
\bibitem[\protect\citeauthoryear{Overzier et al.}{2009}]{overzier2009} Overzier, R.~A., et al., 2009, \apj, 706, 203
\bibitem[\protect\citeauthoryear{Pagel et al.}{1979}]{pagel1979} Pagel, B.~E.~J., Edmunds, M.~G., Blackwell, D.~E., Chun, M.~S., Smith, G., 1979, \mnras, 189, 95
\bibitem[\protect\citeauthoryear{Pellegrini et al.}{2012}]{pellegrini2012} Pellegrini, E.~W., Oey, M.~S., Winkler, P.~F., Points, S.~D., Smith, R.~C., Jaskot, A.~E., Zastrow, J., 2012, \apj, 755, 40
\bibitem[\protect\citeauthoryear{Pentericci et al.}{2011}]{pentericci2011} Pentericci, L., et al., 2011, \apj, 743, 132 
\bibitem[\protect\citeauthoryear{Pettini \& Pagel}{2004}]{PP2004} Pettini, M., Pagel, B.~E.~J., 2004, \mnras, 348, L59
\bibitem[\protect\citeauthoryear{Pettini et al.}{2001}]{pettini2001} Pettini, M., et al., 2001, \apj, 554, 981
\bibitem[\protect\citeauthoryear{Queyrel et al.}{2009}]{queyrel2009} Queyrel, J., et al., 2009, \aap, 506, 681
\bibitem[\protect\citeauthoryear{Rahmati et al.}{2013}]{rahmati2013} Rahmati, A., Schaye, J., Pawlik, A.~H., Rai\v{c}evi\`{c}, M., 2013, \mnras, 431, 2261
\bibitem[\protect\citeauthoryear{Richard et al.}{2011}]{richard2011} Richard, J., Jones, T., Ellis, R., Stark, D.~P., Livermore, R., Swinbank, M., 2011, \mnras, 413, 643
\bibitem[\protect\citeauthoryear{Richardson et al.}{2013}]{richardson2013} Richardson, M.~L.~A., Levesque, E.~M., McLinden, E.~M., Malhotra, S., Rhoads, J.~E., Xia, L., 2013, arXiv e-prints, arXiv:1309.1169
\bibitem[\protect\citeauthoryear{Rigby et al.}{2011}]{rigby2011} Rigby, J.~R., Wuyts, E., Gladders, M.~D., Sharon, K., Becker, G.~D., 2011, \apj, 732, 59
\bibitem[\protect\citeauthoryear{Robertson et al.}{2010}]{robertson2010} Robertson, B.~E., Ellis, R.~S., Dunlop, J.~S., McLure, R.~J., Stark, D.~P., 2010, \nat, 468, 55
\bibitem[\protect\citeauthoryear{Robertson et al.}{2013}]{robertson2013} Robertson, B.~E., et al., 2013, \apj, 768, 71 
\bibitem[\protect\citeauthoryear{Salim et al.}{2007}]{salim2007} Salim, S., et al., 2007, \apjs, 173, 267
\bibitem[\protect\citeauthoryear{Santini et al.}{2014}]{santini2014} Santini, P., et al., 2014, \aap, 562, 30
\bibitem[\protect\citeauthoryear{Savaglio et al.}{2005}]{savaglio2005} Savaglio, S., et al., 2005, \apj, 635, 260
\bibitem[\protect\citeauthoryear{Scarlata et al.}{2009}]{scarlata2009} Scarlata, C., et al., 2009, \apjl, 704, L98
\bibitem[\protect\citeauthoryear{Schenker et al.}{2012}]{schenker2012} Schenker, M.~A., et al., 2012, \apj, 744, 179 
\bibitem[\protect\citeauthoryear{Schenker et al.}{2013}]{schenker2013} Schenker, M.~A., Ellis, R.~S., Konidaris, N.~P., Stark, D.~P., 2013, \apj, 777, 63
\bibitem[\protect\citeauthoryear{Shapley et al.}{2003}]{shapley2003} Shapley, A.~E., Steidel, C.~C., Pettini, M., Adelberger, K.~L., 2003, \apj, 588, 65
\bibitem[\protect\citeauthoryear{Shapley et al.}{2005}]{shapley2005} Shapley, A.~E., Coil, A.~L., Ma, C.~-P., Bundy, K., 2005, \apj, 635, 1006
\bibitem[\protect\citeauthoryear{Shibuya et al.}{2014a}]{shibuya2014a} Shibuya, T., et al., 2014a, arXiv e-prints, arXiv:1402.1168
\bibitem[\protect\citeauthoryear{Shibuya et al.}{2014b}]{shibuya2014b} Shibuya, T., et al., 2014b, \apj, 785, 64
\bibitem[\protect\citeauthoryear{Shields}{1990}]{shields1990} Shields, G.~A., 1990, ARA\&A, 28, 525
\bibitem[\protect\citeauthoryear{Shirazi et al.}{2013}]{shirazi2013} Shirazi, M., Brinchmann, J., Rahmati, A., 2013, arXiv e-prints, arXiv:1307.4758
\bibitem[\protect\citeauthoryear{Stark et al.}{2010}]{stark2010} Stark, D.~P., Ellis, R.~S., Chiu, K., Ouchi, M., Bunker, A., 2010, \mnras, 408, 1628
\bibitem[\protect\citeauthoryear{Stark et al.}{2011}]{stark2011} Stark, D.~P., Ellis, R.~S., Ouchi, M., 2011, \apj, 728, L2
\bibitem[\protect\citeauthoryear{Stark et al.}{2013}]{stark2013} Stark, D.~P., Schenker, M.~A., Ellis, R.~S., Robertson, B.~E., McLure, R., Dunlop, J., 2013, \apj, 763, 129
\bibitem[\protect\citeauthoryear{Steidel et al.}{2010}]{steidel2010} Steidel, C.~C., et al., 2010, \apj, 717, 289
\bibitem[\protect\citeauthoryear{Steidel et al.}{2011}]{steidel2011} Steidel, C.~C., Bogosavljevic, M., Shapley, A.~E., Kollmeier, J.~A., Reddy, N.~A., Erb, D.~K., Pettini, M., 2011, \apj, 736, 160
\bibitem[\protect\citeauthoryear{Storchi-Bergmann et al.}{1994}]{storchi-bergmann1994} Storchi-Bergmann, T., Calzetti, D., Kinney, A.~L., 1994, \apj, 429, 572
\bibitem[\protect\citeauthoryear{Tacconi et al.}{2010}]{tacconi2010} Tacconi, L.~J., et al., 2010, Nature, 463, 781
\bibitem[\protect\citeauthoryear{Tacconi et al.}{2013}]{tacconi2013} Tacconi, L.~J., et al., 2013, \apj, 768, 74
\bibitem[\protect\citeauthoryear{Terlevich et al.}{1993}]{terlevich1993} Terlevich, E., D\'{i}az, A.~I., Terlevich, R., Vargas, M.~L.~G., 1993, \mnras, 260, 3
\bibitem[\protect\citeauthoryear{Tremonti et al.}{2004}]{tremonti2004} Tremonti, C.~A., et al., 2004, \apj, 613, 898
\bibitem[\protect\citeauthoryear{van der Wel et al.}{2011}]{vanderwel2011} van der Wel, A., et al., 2011, \apj, 742, 111
\bibitem[\protect\citeauthoryear{Vanzella et al.}{2006}]{vanzella2006} Vanzella, E., et al., 2006, \aap, 454, 423
\bibitem[\protect\citeauthoryear{Vanzella et al.}{2009}]{vanzella2009} Vanzella, E., et al., 2009, \apj, 695, 1163
\bibitem[\protect\citeauthoryear{Vanzella et al.}{2011}]{vanzella2011} Vanzella, E., et al., 2011, \apjl, 730, L35
\bibitem[\protect\citeauthoryear{Verhamme et al.}{2006}]{verhamme2006} Verhamme, A., Schaerer, D., Maselli, A., 2006, \aap, 460, 397
\bibitem[\protect\citeauthoryear{Verhamme et al.}{2012}]{verhamme2012} Verhamme, A., et al., 2012, \aap, 546, A111
\bibitem[\protect\citeauthoryear{Villar-Mart\'{i}n et al.}{2004}]{villar-martin2004} Viilar-Mart\'{i}n, M., Cervi\~{n}o, M., Delgado, G., 2004, \mnras, 355, 1132
\bibitem[\protect\citeauthoryear{Wuyts et al.}{2012}]{wuyts2012} Wuyts, E., Rigby, J.~R., Sharon, K., Gladders, M.~D., 2012, \apj, 755, 73
\bibitem[\protect\citeauthoryear{Yabe et al.}{2012}]{yabe2012} Yabe, K., et al., 2012, \pasj, 64, 60
\bibitem[\protect\citeauthoryear{Yajima et al.}{2012}]{yajima2012} Yajima, H., Li, Y., Zhu, Q., Abel, T., Gronwall, C., Ciardullo, R., 2012, \apj, 754, 118
\bibitem[\protect\citeauthoryear{Zackrisson et al.}{2013}]{zackrisson2013} Zackrisson, E., Inoue, A.~K., Jensen, H., 2013, \apj, 777, 39
\bibitem[\protect\citeauthoryear{Zahid et al.}{2011}]{zahid2011} Zahid, H.~J., Kewley, L.~J., \& Bresolin, F., 2011, \apj, 730, 137
\bibitem[\protect\citeauthoryear{Zheng \& Wallace}{2013}]{ZW2013} Zheng, Z., Wallace, J., 2013, arXiv e-prints, arXiv:1308.1405
\end{thebibliography}
\end{document}